\documentclass[twocolumn,twocolappendix]{aastex631}

\usepackage{listings}
\usepackage{color} 

\definecolor{codegreen}{rgb}{0,0.6,0}
\definecolor{codegray}{rgb}{0.5,0.5,0.5}
\definecolor{codepurple}{rgb}{0.58,0,0.82}
\definecolor{backcolour}{rgb}{0.95,0.95,0.92}

\shorttitle{TransformerPayne: precise and data-efficient emulation}

\begin{document}

\title{TransformerPayne: enhancing spectral emulation accuracy and data efficiency by capturing long-range correlations}

\correspondingauthor{Tomasz R\'o\.za\'nski}
\email{Tomasz.Rozanski1@anu.edu.au}

\author[0000-0002-5819-3023]{Tomasz R\'o\.za\'nski}
\affiliation{Research School of Astronomy \& Astrophysics,\\ 
             The Australian National University,\\
             Cotter Rd., Weston, ACT 2611, Australia}
\affiliation{Astronomical Institute,\\
             University of Wroc\l aw,\\
             Kopernika 11, 51-622 Wroc\l aw, Poland}

\author[0000-0001-5082-9536]{Yuan-Sen Ting}
\affiliation{Department of Astronomy,\\
             The Ohio State University,\\ 
             Columbus, OH 45701, USA}
\affiliation{Center for Cosmology and AstroParticle Physics (CCAPP),\\ 
             The Ohio State University,\\ 
             Columbus, OH 43210, USA}

\author[0000-0001-6962-4979]{Maja Jab\l o\'nska}
\affiliation{Research School of Astronomy \& Astrophysics,\\ 
             The Australian National University,\\
             Cotter Rd., Weston, ACT 2611, Australia}

\begin{abstract}

Stellar spectra emulators often rely on large grids and tend to reach a plateau in emulation accuracy, leading to significant systematic errors when inferring stellar properties. Our study explores the use of Transformer models to capture long-range information in spectra, comparing their performance to The Payne emulator (a fully connected multilayer perceptron), an expanded version of The Payne, and a convolutional-based emulator. We tested these models on synthetic spectra grids, evaluating their performance by analyzing emulation residuals and assessing the quality of spectral parameter inference. The newly introduced TransformerPayne emulator outperformed all other tested models, achieving a mean absolute error (MAE) of approximately 0.15\% when trained on the full grid. The most significant improvements were observed in grids containing between 1000 and 10\,000 spectra, with TransformerPayne showing 2 to 5 times better performance than the scaled-up version of The Payne. Additionally, TransformerPayne demonstrated superior fine-tuning capabilities, allowing for pretraining on one spectral model grid before transferring to another. This fine-tuning approach enabled up to a tenfold reduction in training grid size compared to models trained from scratch. Analysis of TransformerPayne's attention maps revealed that they encode interpretable features common across many spectral lines of chosen elements. While scaling up The Payne to a larger network reduced its MAE from 1.2\% to 0.3\% when trained on the full dataset, TransformerPayne consistently achieved the lowest MAE across all tests. The inductive biases of the TransformerPayne emulator enhance accuracy, data efficiency, and interpretability for spectral emulation compared to existing methods.

\end{abstract}

\keywords{Stellar atmospheres(1584) --- Galactic archaeology(2178) --- Astroinformatics(78) --- Astrostatistics(1882)}

\section{Introduction}

Spectroscopy is a cornerstone in astrophysics, providing the key to understanding the complex evolution and properties of stars, galaxies, and other astrophysical objects and phenomena. The field of stellar spectroscopy has seen considerable evolution due to large-scale surveys like APOGEE, LAMOST, Gaia-ESO, and GALAH \citep{Gilmore2012,Luo2015,Majewski2017,Buder2020}, which calls for a better analysis techniques to manage the influx of high-quality spectral data. The shift from analyzing a few thousand spectra \citep{Fuhrmann1998,Bensby2003} to handling millions is illustrated by the upcoming 4MOST survey \citep{2019Msngr.175....3D}, which will acquire 20 million low-resolution (R $\approx$ 6500) and 3 million medium-resolution (R $\approx$ 20\,000) spectra over five years of operation. Similarly, the WEAVE survey \citep{2014SPIE.9147E..0LD} projects comparable figures. The scale of these surveys required advancements in stellar spectra modeling, that led to transitioning from detailed star-by-star analyses using spectral synthesis codes to pipelines that employ efficient amortization techniques, such as neural network emulation and inference tools, capable of processing this volume of stellar spectra.

An important aspect of building a pipeline that provides reliable estimates of parameters of stellar atmospheres is access to physics-based numerical stellar atmospheric models that include all relevant physical phenomena as comprehensively as possible. There are many tools generally available for inferring stellar parameters. To name just a few, there are SME \citep{2017AA...597A..16P}, iSpec \citep{2014AA...569A.111B}, FAMA \citep{2013AA...558A..38M} or GALA \citep{2013ApJ...766...78M}, which rely on spectrum synthesis numerical codes or interpolation across extensive grids. At the forefront of today's research, non-local thermodynamic equilibrium (non-LTE) and 3D effects are of the greatest importance when targeting precise stellar atmospheric parameters, such as effective temperature and surface gravity or elemental abundances \citep{Magg2022, 2022AA...668A..68A, 2023AA...677A..98Z}. There are still many areas to advance, including extending modeling toward shorter \citep{2020Galax...8...60H} and longer wavelengths \citep{2022AA...666A..62L}, incorporating non-LTE effects for more lines and atoms through the development of more comprehensive atomic models \citep{2010EAS....43..115P}, considering the vertical stratification of elements in stellar atmospheres \citep{2009AA...495..937L} and including the influence of magnetic fields \citep{2024AA...684A.175H}. Additionally, extending modeling toward lower temperatures, where molecular lines and possibly even weather might be of interest, and toward higher temperatures, where detailed modeling of winds is crucial, is also important.

Inference of atmospheric parameters, whether through optimization  (e.g., mean-squared error) or posterior sampling, involves thousands of evaluations of spectrum synthesis code and, ideally, atmospheric structure calculations. This makes it infeasible when considering the state-of-the-art models, which can take hours to days to converge. This requires the development of amortization methods, which involve running large-scale initial calculations to enable later fast and accurate inference.

Traditionally this is handled by the computation of large spectral grids covering the parameter space of interest and later using interpolation, and often some additional post-processing, e.g., convolution with a rotational kernel, in the inference part. This approach quickly becomes infeasible as the size of the grid necessary for interpolation grows exponentially with the number of inferred parameters, especially when we are targeting the inference of dozens of individual abundances it is no longer within current computational reach.

A solution to this \textit{dimensionality curse} is to replace interpolation using traditional methods, with modeling-based approach, referred to as emulation. Models used in this context no longer perform bare interpolation but are optimized to approximate the complex function from spectrum parameters to normalized fluxes. Pioneering works in this domain use quadratic (e.g., The Cannon by \citet{2015ApJ...808...16N}), and polynomial modeling \citep{2016ApJ...826L..25R}. These approaches, however, might not be expressive enough to precisely model the complexities of stellar spectra, such as the highly non-linear and correlated behavior of many atomic features.

The neural network based approach, which has been empirically proven to be very flexible and efficient in high dimensional emulation with sparse data, i.e. with a number of spectra smaller than exponential in the number of dimensions. The first proposed neural network architecture applied to this task is The Payne model, which is a multi-layer perceptron (MLP; dense network; The Payne in the context of astrophysics \citep{2019ApJ...879...69T}). It is a simple and robust model, offering fast optimization (in the context of neural networks called \textit{training}) and high prediction accuracy. Despite its performance, the main limitation of The Payne model is its saturation when the mean absolute error of emulation is around 0.01 in normalized flux. Precise error depends on the dimensionality and the span of the grid, but the 0.01 reported here is illustrative of the expected order of magnitude for the emulation error of The Payne.

Saturation at this level hinders the usage of this emulator when modeling effects that weakly manifest in spectra. For example, the inclusion of non-LTE effects in stellar spectra calculation affects only a small subset of spectral lines that are present in stellar spectra. The change introduced by this additional physics is small and in many cases can be of the order of several percent in normalized flux. When relating the strength of this signal, to the mean absolute accuracy of 0.01 in normalized flux for which The Payne tends to saturate, it becomes evident that the usage of this emulator might hinder the benefits of having much more complex physical models. Other effects that mostly influence the shape of spectral lines, like hyperfine splitting, differential rotation, pulsations, starspots or Zeeman splitting, also manifest themselves in weak signals, so accurate emulation is necessary for survey-scale inference of these effects. Additionally, these detailed physical effects are often computationally expensive, allowing only small grids of hundreds to thousands of spectra to be feasible.

The paper is structured as follows: Section\,\ref{sec:relevant_ML} discusses the key motivation of this paper and explains the concept of inductive bias, Section\,\ref{sec:transformer_payne} describes the methods, introduces the TransformerPayne architecture and outlines the details of data and training. Experiments are described in Section\,\ref{sec:results} and discussed in Section\,\ref{sec:discussion}. The conclusions and future work are detailed in Section\,\ref{sec:conclusions}.

\begin{figure*}
\centering
\includegraphics[width=1.0\hsize]{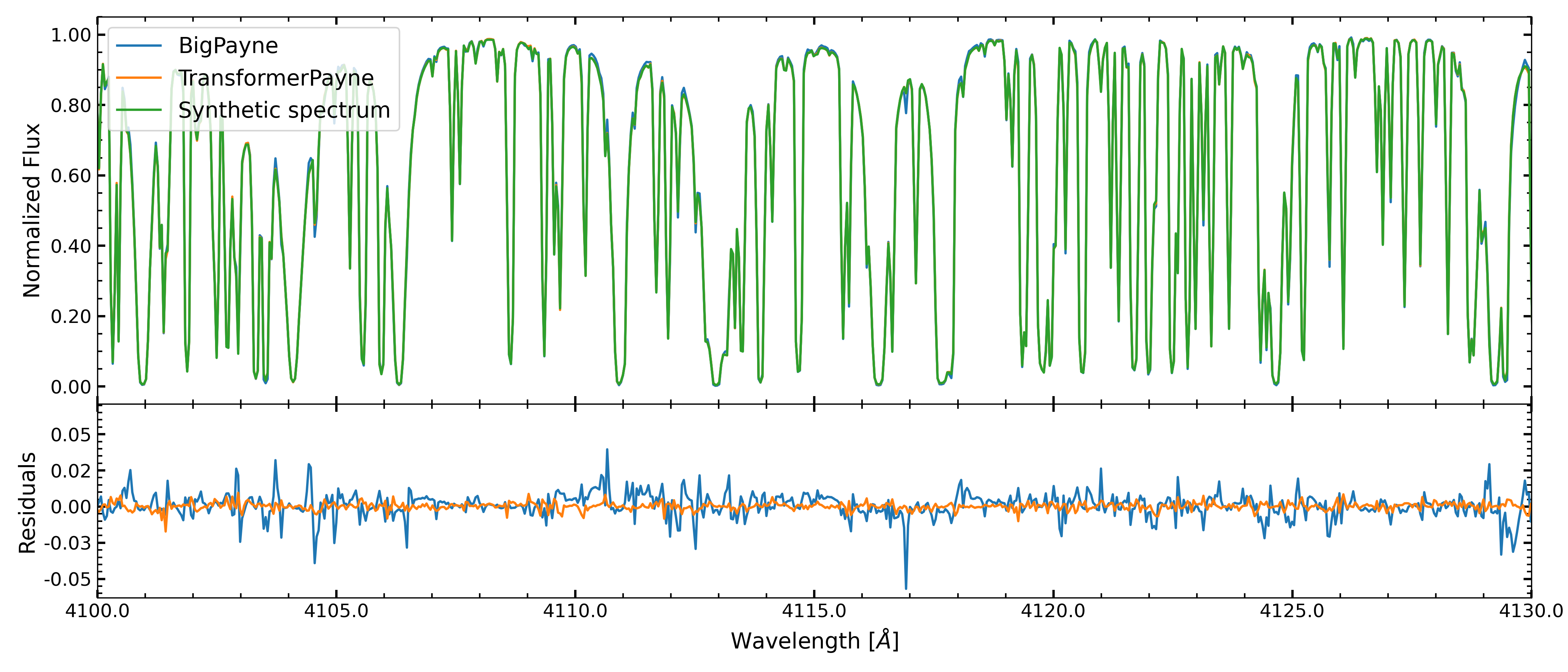}
    \caption{Illustrative predictions and residuals comparing of the TransformerPayne in this study and BigPayne model, which is a scaled up The Payne emulator, both trained on the training grid from scratch, with the same training set of 100\,000 spectra. The residuals of the TransformerPayne are reduced when compared to BigPayne residuals, demonstrating that the strong inductive biases of TransformerPayne can lead to more precise emulation of spectra.}
    \label{fig:fig1_residuals_4100_4130}
\end{figure*}

\section{Inductive bias}
\label{sec:relevant_ML}

One of the limiting factor of using the simple multi-layer perceptron models (The Payne) model to emulate spectra stems from the fact of its inadequate \textit{inductive bias}. Inductive biases can be defined as a set of rules encoded in the machine learning model or learning process, which are used when predicting output for an unseen input. Inductive bias allows the model to choose one particular solution from others, even when they give the same results. The \textit{Occam's razor} rule, which prioritizes the simplest solutions, is a classical example of inductive bias. Inductive biases of neural networks mostly manifest in the choice of model's architectures. The multi-layer perceptron architecture assumes one of the weakest inductive biases among neural network architecture. This architecture assumes only the smoothness of the approximated function and does not have any inherent mechanisms that exploit local or sparse structure of the approximated function. As The Payne is generally a relatively small MLP which, due to its moderate expressiveness, still can be considered to enforce a simplicity bias.

Inductive biases not fitted for the precise stellar spectrum emulation are probably the most important factor causing saturation of The Payne architecture-based emulators. Investigation of the influence of simplicity bias can be tested by scaling The Payne to significantly bigger networks, which we will also explore in this study. But more importantly, this study aims to explore inclusion of inductive biases associated with the local and long-distance structure of stellar spectra by testing other neural network architectures. Such inductive biases might lead to better emulation of spectra. On the on hand, local structure is associated with the significant width of spectral features and profiles, with hydrogen lines or molecular bands as wide as 100\,\r{A}. On the other hand, there is a sparse and complex structure across different lines of the same elements. In particular, there might be a positive correlation for all lines of a given element if its abundance increases, or a more complex dependencies in the case of temperature variation.

An architecture that scales well to large networks and captures complex dependencies between various parts of the input is the Transformer architecture \citep{2017arXiv170603762V}. First developed for natural language processing, it was later adopted for image, audio, and video processing, is gradually becoming a current one-fits-all solution in machine learning. Its adoption extends much beyond the mentioned modalities. In particular, TimesLM is a general model for time series processing \mbox{\citep{2023arXiv231010688D}}, Genomic Pre-trained Network is a model trained on genomic DNA sequences \citep{Benegas2022}, and ProteinLM is a model trained on protein sequences \citep{2021arXiv210807435X}.

The appropriate choice of inductive biases is critical to overcome the issue of model saturation and can lead to better data efficiency, i.e., in the context of stellar spectra emulation, means that a fixed targeted emulation uncertainty can be achieved with a smaller number of spectra. A high data-efficiency is particularly important for spectral modeling, as proper numerical modeling of the stellar atmosphere and the emergent spectrum, which includes complex physics such as 3D-NLTE models, can be computationally expensive, and calculation of extensive grids might be infeasible. 

\section{Methods and the TransformerPayne architecture}
\label{sec:transformer_payne}

The objective of this study was to construct a stellar spectrum emulator capable of accurately replacing a numerical model in parameter estimation pipelines. This emulator approximates a function, $f(\lambda,\vec{p})$, where $\lambda$ represents the wavelength, and $\vec{p}$ denotes a set of arbitrary input parameters, such as effective temperature or elemental abundances. For comparison, we evaluated three different neural networks. 

As the baseline, we utilized a basic approach known as the Multilayer Perceptron (MLP), also referred to as The Payne in astronomical literature \citep{2019ApJ...879...69T, Straumit2022, Xiang2022}\footnote{We based our implementation on the one available at \url{https://github.com/tingyuansen/The_Payne}.}. This method predicts fluxes at a fixed set of wavelengths. MLP is the main building block of most machine learning models and is a simple vector-matrix multiplication with nonlinear element-wise function. We used network of the size typically adopted for stellar spectrum emulation and comparable to the original The Payne, a three-layer MLP:
\begin{equation}
    \vec{f} = \mathbf{W_3}\textrm{gelu}(\mathbf{W_2}\textrm{gelu}(\mathbf{W_1}\vec{p} + \vec{b}_1) + \vec{b}_2) + \vec{b}_3,
\end{equation}
where $\vec{f}$ is a predicted vector of normalized fluxes, matrices $\mathbf{W_i}$ and vectors $\vec{b_i}$ are free parameters of the model and $\vec{p}$ is the vector of input parameters. The shapes of matrices $\mathbf{W_i}$ are respectively $p_{\text{dim}} \times 128$, $128 \times 128$ and $128 \times 22315$. Alternatively the architecture can be described as having 3 layers with 128, 128 and 22315 neurons in subsequent layers. Lastly, $\text{gelu}(x)$ is an element-wise non-linearity function \citep[][in the context of machine learning, non-linearities are referred to as activation functions]{2016arXiv160608415H}. The $\textrm{gelu}(x)$ function is:
\begin{equation}
\textrm{gelu}(x) = x \cdot \frac{1}{2} \left(1 + \text{erf}\left(\frac{x}{\sqrt{2}}\right)\right),
\end{equation}
where $\text{erf}(x)$, known as the error function, is defined by an integral:
\begin{equation}
\text{erf}(x) = \frac{2}{\sqrt{\pi}} \int_0^x e^{-t^2} \, dt.
\end{equation}
Additionally to The Payne model, we explored scaling it to a larger network, and subsequently we call this version the BigPayne. It has 4 layers with 2048, 2048, 2048 and 22315 neurons.

We also contrast our study with the Convolutional-based neural network \citep{LeCun1989,NIPS2012_c399862d, Zeiler2013}, which are also often used for the processing of astrophysical spectra, e.g., in the context of classification of stellar spectra \citep{2020MNRAS.491.2280S} or estimation of quasar redshifts \citep{2022MNRAS.511.4490R}. In the Convolutional-based emulator, the MLP part is used to predict a low-resolution spectrum embedding, followed by several convolutional layers that aim to learn up-sampling function. MLP part has three layers with 2048 neurons in each layer, then the result of MLP is reshaped to $512 \times 4$ matrix which can thought as the low resolution spectrum tokenization. Then this is processed using five up-sampling blocks which outputs resolution of 16384, equals $512\times 2^5$, later linearly interpolated to the normalized flux output size of 22135. This architecture is motivated by the idea of separating long-distance correlations with the MLP and resolving short-term interactions (e.g., line shapes) using learned up-sampling blocks.

The parameter count for all considered emulators is approximately: 3M for The Payne, 54M for BigPayne, 9M for Convolution-based, and 17M for TransformerPayne. We do not compare to more traditional interpolation methods as they are infeasible for high dimensional grids explored in this work, and their weaknesses are discussed in detail in \cite{2019ApJ...879...69T}. Details of the architectures can be found in the code listings in the Appendix\,\ref{arch_details}. 

\subsection{TransformerPayne architecture}
\label{sec:TP_architecture}

The Transformer architecture \citep{2017arXiv170603762V} has shown effectiveness in multiple domains, but various modifications are critical to adapt to the specificities of different data. In particular, spectral data are distinct from modalities like video or audio as flux is a function of wavelength without shift invariance. Because of this characteristic, we propose a variation of the Transformer architecture which explicitly depends on wavelength. More specifically, our model will predict  $\text{Normalized Flux} = f(\lambda, \vec{p})$ where $\vec{p}$ is a vector of stellar atmospheric parameters and $\lambda$ is a single wavelength. From the early layers of the architecture, parametrization in wavelengths biases the model to learn features shared across lines of the same element, even if they are widely separated in wavelength. This is achieved through the flexibility of the attention mechanism, the main building block of the Transformer architecture (see detailed description below). During inference, the emulator can be vectorized to predict the Normalized Flux at the arbitrarily chosen wavelength grid. This simplifies the modeling of spectra from an arbitrary instrument and also makes it simple to study the effects of Doppler shifts without additional steps of interpolation.

The TransformerPayne model is implemented using usual building blocks of Transformer-based architecture and builds on the work on conditioning on spatial dimensions in the field of Neural Radiance Fields \citep{2020arXiv200308934M, 2021arXiv211113152S, 2022arXiv220910684R}. Block-by-block description of the architecture follows below.

\begin{figure}
\centering
\includegraphics[width=0.85\hsize]{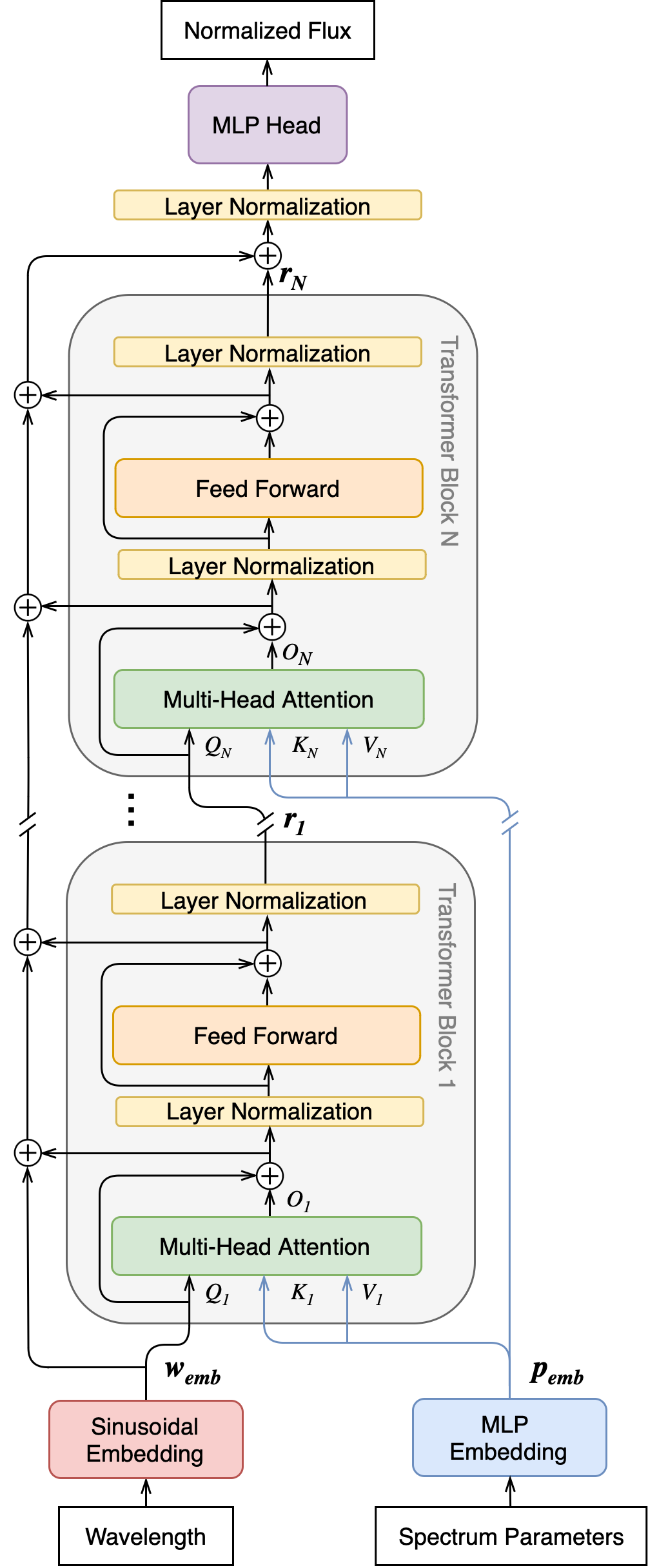}
  \caption{Architecture of the TransformerPayne stellar spectra emulator: The model has two inputs: wavelength and a vector of spectrum parameters. Wavelength is encoded into a query vector via sinusoidal encoding, while the parameters are transformed into sequence of vectors using an MLP Embedding. Transformer Blocks capture long-range information, and the normalized flux is predicted using an MLP Head.}
     \label{fig:fig2_architecture}
\end{figure}

\subsubsection*{Wavelength and Spectrum Parameters embedding modules}

Two inputs to the TransformerPayne emulator, the wavelength (a scalar value) and a vector of spectrum parameters (e.g., effective temperature, surface gravity, and individual abundances), are first fed into corresponding embedding modules. In this context, \textit{embedding} refers to a transformation into a domain that improves efficiency of the latter components of TransformerPayne, specifically the Transformer Blocks. These blocks are effective at modeling complex dependencies between sequences of high-dimensional vectors, often referred to as \textit{tokens}, in the context of machine learning. Based on initial tests we employed 256-dimensional ($d=256$) tokens throughout the entire architecture.

To embed a scalar value of a wavelength into a high-dimensional vector space, we employed the Sinusoidal Embedding which computes the function:
\begin{equation}
\vec{w_{emb}} = \big(\sin (\omega_1 w), \sin (\omega_2 w), \ldots, \sin (\omega_d w)\big),
\end{equation}
where $\omega_1$, ..., $\omega_d$ is a sequence of angular frequencies chosen to cover the wavelength span of characteristic spectral features. As a rule of thumb, the smallest angular frequency should correspond to the range between the minimum and maximum wavelengths, while the highest should correspond to the scale of narrow absorption features or the resolution of the targeted spectral grid. In this work we used a decimal logarithm of wavelength as a wavelength coordinate, and we parameterized the vector of $\omega_i = 2\pi / P_i$ using 256 equidistantly spaced periods $P_i$, ranging from $P_{min}=10^{-6}$ to $P_{max} = 10$.

While the wavelength is embedded into a sequence consisting of a single vector, the spectral parameters vector is embedded into a sequence of 16 tokens ($t = 16$). This embedding is performed using an MLP Embedding, which is a simple two-layer perceptron followed by the reshaping of the output vector into a matrix with dimensions $16 \times 256$. The function computed by the MLP Embedding is:
\begin{equation}
\vec{p_{emb}} = \mathbf{W}_2~\textrm{gelu}\big(\mathbf{W}_1 \vec{p} + \vec{b_1}\big) + \vec{b_2},
\end{equation}
where $\vec{p}$ represents a vector of spectral parameters, $\mathbf{W}_1$ and $\mathbf{W}_2$ are weight matrices with shapes $p_{\text{dim}} \times 1024$ and $1024 \times 4096$ respectively, $\vec{b}_1$ and $\vec{b}_2$ are bias vectors, and $\text{gelu}(x)$ is an element-wise non-linearity function.

The length of the embedding sequence is chosen based on our precursor experiments, but it might be further optimized in the future. Generally, a smaller number of tokens results in faster model performance, while an increase in the length of the embedding sequence is associated with improved prediction accuracy.

\subsubsection*{Transformer Block}

Upon the tokenization step, the tokens are then passed through the Transformer Block, which consists of a Multi-Head Attention and a Feed Forward modules. In the tested TransformerPayne, Transformer Block is repeated 16 times ($N=16$).

\begin{figure}
\centering
\includegraphics[width=0.85\hsize]{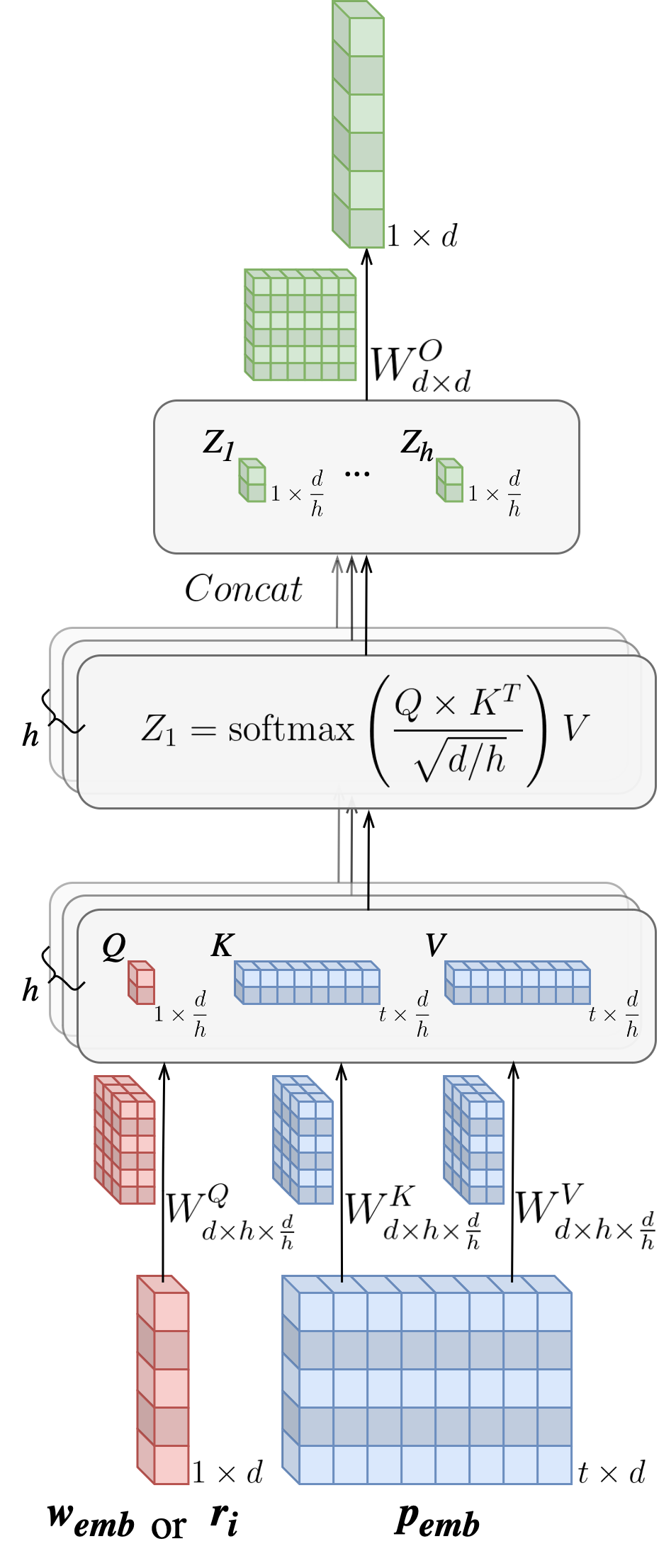}
  \caption{Illustration of Multi-Head Attention, building block of TransformerPayne emulator. It enables the emulator to learn the conditioning of the embedding of wavelength $\vec{w_{emb}}$ or the output from the previous transformer block $\vec{r_i}$ on the embedding of stellar spectra parameters $\vec{p_{emb}}$. The dot-product weighted attention function is the central element of this module, which enables the capture of long-range dependencies due to similarities between embeddings as captured by the product of Query and Key matrices ($\mathbf{Q}_i \times \mathbf{K}^T_i$).
}
     \label{fig:fig3_MHA}
\end{figure}

The Multi-Head Attention (MHA) block is responsible for fine-grained conditioning of its output on wavelength and spectrum parameters, as detailed in Fig.\,\ref{fig:fig3_MHA}. It has three inputs: Query (Q), Key (K), and Value (V). In the first Transformer Block, the Query input is the sinusoidal embedding of the wavelength ($\vec{w_{emb}}$), while in subsequent Transformer Blocks, this input takes the output from the previous Transformer Block ($\vec{r}_i$). The input to both Key and Value is shared across all Transformer Blocks and consists of an embedding of spectral parameters ($\vec{p_{emb}}$). For a clear illustration, see Fig.\,\ref{fig:fig2_architecture}, which depicts the architecture of the TransformerPayne emulator.

The function that Multi-Head Attention calculates is described in detail below. First, it linearly transforms its inputs into $\mathbf{Q}$, $\mathbf{K}$ and $\mathbf{V}$ matrices:
\begin{equation}
\mathbf{Q} = W^{\mathbf{Q}} \vec{x} + \vec{b^{\mathbf{Q}}},
\end{equation}
\begin{equation}
\mathbf{K} = W^{\mathbf{K}} \vec{p_{emb}} + \vec{b^{\mathbf{K}}},
\end{equation}
\begin{equation}
\mathbf{V} = W^{\mathbf{V}} \vec{p_{emb}} + \vec{b^{\mathbf{V}}},
\end{equation}
where $\vec{x}$, fed into the Query input, is the embedding of wavelength, $\vec{p_{emb}}$, in the case of the first transformer block, or an output from the previous Transformer Block, $\vec{r_i}$, in the case of the rest of Transformer Blocks. This linear transformation can be thought of as a transformation to a domain where the dot product, $\mathbf{Q} \times \mathbf{K}^T$, quantifies contextual similarity between tokens, and the Value embedding is modified to extract the content relevant for subsequent weighted dot-product attention. Multi-Head Attention, which we are using, reshapes $\mathbf{Q}$, $\mathbf{K}$, and $\mathbf{V}$ into distinct \textit{heads}. These heads are used in $h$ independent dot-product attention operations, yielding outputs $Z_1$ to $Z_h$ ($h=8$). Dot-product weighted attention, which facilitates the conditioning, is given by:

\begin{equation}
Z_i = \mathbf{A}_i \mathbf{V}_i = \text{softmax}\Bigg(\frac{\mathbf{Q}_i \times \mathbf{K}^T_i}{\sqrt{d/h}}\Bigg) \mathbf{V}_i,
\end{equation}
where $\text{softmax}(x)$ is:
\begin{equation}
\text{softmax}(\vec{x}) = \frac{1}{\sum_{j=1}^t e^{x_j}} (e^{x_1}, e^{x_2}, \dots, e^{x_t}),
\end{equation}
which turns rows of unnormalized attention matrix, $\frac{\mathbf{Q}_i \times \mathbf{K}^T_i}{\sqrt{d/h}}$, into discrete probabilities over the sequence of tokens, $\mathbf{V}_i$. Note that in our case, the attention matrix, $\mathbf{A}_i$, is a one-row matrix as the Query input is always a one-token sequence.

Each \textit{head} returns a vector $Z_i$ and is supposed to learn to attend to different parts of the spectrum parameter embedding, $\vec{p_{emb}}$. This enables the Multi-Head Attention block to learn multi-turn conditioning of the predicted normalized flux on all relevant parameters, across a wide range of wavelengths, dealing with long-span correlations in stellar spectra. In the next step, all vectors are concatenated in vector $\vec{z}$ and linearly processed to produce the MHA block output:
\begin{equation}
\vec{y} = \mathbf{W^\mathbf{O}} \vec{z} + \vec{b}^{\mathbf{O}}.
\end{equation}

The second module of Transformer Block, which is a Feed Forward neural network, modifies an output of MHA block using a simple two-layer perceptron:

\begin{equation}
\vec{r} = \mathbf{W}_2~\textrm{gelu}\big(\mathbf{W}_1 \vec{y} + \vec{b_1}\big) + \vec{b_2},
\end{equation}
where $\vec{y}$ is a $256$-dimensional input vector, $\mathbf{W}_1$ and $\mathbf{W}_2$ are weight matrices (with corresponding shapes $256\times 1024$ and $1024 \times 256$), $\vec{b_1}$ and $\vec{b_2}$ are bias vectors.

\subsubsection*{MLP Head Block}

The last module of the TransformerPayne architecture is the MLP Head Block. It is responsible for predicting the Normalized Flux from a 256-dimensional representation built by the preceding 16 Transformer Blocks and is implemented as a two-layer perceptron:
\begin{equation}
\text{f} = \mathbf{W}_2~\textrm{gelu}\big(\mathbf{W}_1 \vec{r} + \vec{b_1}\big) + \vec{b_2},
\end{equation}
where f is predicted Normalized Flux, $\vec{r}$ is a $256$-dimensional input vector, $\mathbf{W}_1$ and $\mathbf{W}_2$ are weight matrices (with corresponding shapes $256\times 256$ and $256 \times 1$), $\vec{b_1}$ and $\vec{b_2}$ are bias vectors.

\subsubsection*{Placement of residual connections and normalization layers}

Stability of training of deep neural networks depends on the gradient landscape of employed loss function. Vanishing and exploding gradients are the typically encountered issues. Both are associated with exponential effect of sequential application of trainable neural network blocks, in our case Transformer Blocks. Among the most important methods to overcome these instabilities is usage of residual-connections and normalization layers.

Residual connections shortens the path of the gradient by usage of skip-connections that parameterize building blocks of neural network as:
\begin{equation}
\vec{y} = \text{NeuralNetworkBlock}(\vec{x}) + \vec{x}.
\end{equation}
Residuals connection helps to initialize neural network blocks close to identity function, which means that both the values and gradients are initially passed unchanged to very deep layers of considered neural network, which facilitates stable training. The blocks of TransformerPayne that use residual connections are the Multi-Head Attention and Feed Forward blocks in all TransformerBlocks.

Normalization layers are the second important component that prevents the exploding gradient problem and speeds up training. They prevent exploding gradients by constraining the output to a chosen distribution, which translates to constrained gradients. In this work we used LayerNormalization, $LN(\vec{x})$ \citep{2016arXiv160706450L} which normalizes its inputs, $\vec{x}$:
\begin{equation}
LN(\vec{x}) = \left(\frac{\vec{x} - \mu}{\sigma}\right) \vec{\alpha} + \vec{\beta},
\end{equation}
where $\mu$ and $\sigma$ are mean and standard deviation of $\vec{x}$, and $\vec{\alpha}$ and $\vec{\beta}$ are trainable vectors. 

In the TransformerPayne model, the placement of residual connections and normalization layers mostly follows the recommendations from the work by \citet{2023arXiv230414802X}, known as ResiDual. ResiDual addresses the issues found in the two most common schemes: Post-Layer Normalization (Post-LN) and Pre-Layer Normalization (Pre-LN). Post-LN suffers from the gradient vanishing problem, while Pre-LN experiences representation collapse, where representations in deep layers of Transformer-based architectures become very similar. This similarity reduces the model's capacity which harms a model accuracy. Placement of residuals connections and normalization layers in TrasformerPayne is illustrated in the Fig.\,\ref{fig:fig2_architecture} and in detail in the Appendix\,\ref{arch_details} in the code Listing\,\ref{lst:transformer_payne}.

\subsection{Training data}

Experiments on stellar spectra emulators and transfer learning require large grids of spectra to investigate how training set size affects the emulation precision. This motivates the calculation of synthetic spectra using Local Thermodynamic Equilibrium (LTE) approximation in all conducted experiments, as LTE codes are efficient and robust. In this study, we will also explore the pre-training -- fine-tuning scenario, where we will train on one domain, and then fine-tune the model on the other domain, as such we also requires distinct domains. Pre-training is conducted in a domain where temperature and surface gravity are fixed, while the targeted domain contains spectra with those parameters randomly sampled from chosen range.

Both grids of synthetic spectra, are produced using updated plane-parallel atmospheric model codes as revised by \citet{2008A&A...491..633L}. These revisions build upon the standard LTE models ATLAS/SYNTHE as detailed by \citep{1979ApJS...40....1K, 1993KurCD..18.....K,Kurucz2005,Kurucz2013}. Each of these sets comprises 100\,000 spectra, generated at a resolution of $R=100\,000$ and covering a wavelength range from 4000 to 5000\,\r{A}. Both spectra grids have a microturbulence velocity of $\xi=0$\,km/s, and a helium content that vary from none to twice the solar value (number fraction from 0.0 to 0.1564). 

The abundance of all other elements with atomic number between 3 (Li) and 99 (Es) is distributed uniformly and independently within a range of $-2$ to 1\,dex relative to solar abundance. We decided to vary all individual elements available in ATLAS/SYNTHE and train emulators with all input abundances, even if lines of a given element are not present in the considered wavelength. This approach enables more in depth analysis of how the dependence of output imprints on the function learned by emulators, especially in the more complex TransformerPayne architecture.

The pre-training grid was defined with a constant effective temperature of $5000$\,K, a logarithm of surface gravity of $\log g=4.5$. In constructing the fine-tuning grid, the effective temperature varied from 4000\,K to 6000\,K and the logarithm of surface gravity from 4.0 to 5.0, while maintaining the same conditions for the other parameters as in the pre-training grid. Details of covered parameters can be found in the Table\,\ref{tab:grids}.

We want to emphasize that usage of LTE modeling is a useful simplification that does not affect our main scientific goals. We aimed to evaluate various neural emulator architectures, including the developed TransformerPayne architecture, focusing on the impact of training data volume, scalability, and the feasibility of applying transfer learning. The scenario of transfer learning considered here has direct parallels to the application of transfer learning from the domain of 1D LTE models to 3D non-LTE, where additional physics affect the shapes and strengths of lines but not their general presence. Fine-tuning between grids of stellar spectra with different line lists presents additional complexities, which, while interesting, are left for future work.

\begin{table*}
\centering
\caption{Grids of synthetic spectra adopted as training set in this study.}
\label{tab:grids}
\begin{tabular}{lll}
\hline
Grid definition & Pre-training & Training  \\
\hline
Effective Temperature & 5000 K & $[4000,6000]$ K \\
Surface Gravity & 4.5 & $[4.0,5.0]$ \\
\# Training Spectra & \multicolumn{2}{c}{Up to 
 100\,000} \\
Wavelength range & \multicolumn{2}{c}{$[4000,5000]$\,\r{A}} \\
\# Wavelengths & \multicolumn{2}{c}{22315} \\
Microturbulence, $\xi$ & \multicolumn{2}{c}{0 km/s} \\
Helium Abundance & \multicolumn{2}{c}{$[0,0.1568]$} \\
Other Abundances, ${\rm [X/H]}^*$ & \multicolumn{2}{c}{$[-2,1]$} \\
\hline
\end{tabular}
\\[5pt]
\textit{Note: $^*$Other Abundances include elements with atomic numbers $Z$ from 3 (Li) to 99 (Es), totaling 97 individual abundances of metals ($Z>2$).}
\end{table*}

\subsection{Training and metrics}

The optimization of large neural networks, conventionally referred to as training, is based on iterative updates of model's free parameters using chosen optimization algorithm. What is optimized is empirical loss function, defined as a discrepancy between model's predictions and the base truth over a set of examples. Most optimization algorithms rely on the gradient of loss function with respect to all (often millions or billions) free parameters of the trained neural network. The gradient is efficiently obtained using the back-propagation algorithm \citep{Rumelhart1986LearningRB}. Each training step includes the creation of a batch of input-output pairs, the evaluation of the loss function and its gradients with respect to all free parameters of the neural network, the update of the optimization algorithm that returns the corrections to be applied to the weights, and the application of these corrections. At the beginning, the dataset is usually split into two parts: the training set, from which the batches of input-output pairs are sampled and used in gradient updates, and the validation set, which is used to monitor if the model is generalizing to unseen data.

In this work all experiments utilized a consistent training strategy. We employed the AdamW optimizer \citep{2014arXiv1412.6980K, 2017arXiv171105101L} with a peak learning rate of $10^{-4}$, a batch size of 16, and a cosine rate scheduler with a linear warm-up for the first 10\% of steps. This means that the learning rate is first initialized to zero, then it grows linearly for the first 10\% of training steps to the maximum learning rate, and then decay to zero following a template shape of cosine function on the domain from 0 to $\pi$. Unless otherwise indicated, the training was conducted for $10^6$ training steps, applicable both to training from scratch and to fine-tuning experiments. We trained our models by minimizing the Mean Squared Error (MSE) across a set of training pairs $\{(\vec{\lambda}, \vec{p})_i, \vec{y}_i\}_N$:
\begin{equation}
\text{MSE}(\{(\vec{\lambda}, \vec{p})_i, \vec{y}_i\}_N) = \frac{1}{N} \sum_{i=1}^{N} \frac{1}{M} \sum_{j=1}^{M} \left( y_{ij} - f(\lambda_{ij},\vec{p}_i) \right)^2,
\end{equation}
where $N$ is number of spectra in a batch, $M$ is the number of wavelengths in a spectrum, $\lambda_{ij}$ is a wavelength, $\vec{p}_i$ is a vector of stellar spectrum parameters, and $y_{ij}$ is the normalized flux for given $(\lambda, \vec{p})_i$. In all models, except the TransformerPayne, $\vec{\lambda}_{i} = \{\lambda_j\}_i$ are shared for all spectra and contain 22315 wavelengths from original synthetic grids. For the TransformerPayne model, which explicitly parameterize output on the vector of wavelengths, 8192 wavelengths were uniformly sampled for every example in the batch and normalized fluxes were linearly interpolated on those wavelengths. Please note that, as TransformerPayne uses fewer wavelengths in each training step, $10^6$ steps correspond to 160 epochs for The Payne, BigPayne, and Convolutional-based models, and to about 60 epochs for the TransformerPayne model. Nonetheless, we consider the comparison fair as the number of gradient updates is consistent across the models, and the results of spectrum interpolation with 22315 and 8192 samples are comparable, returning close gradient estimates over the batch.

All reported metrics were calculated using a validation dataset comprised of 1024 spectra. These spectra were sampled from the same domain as the training grid. For our metrics, in addition to the Mean Squared Error, we utilized Mean Absolute Error (MAE):
\begin{equation}
\text{MAE}(\{(\vec{\lambda}, \vec{p})_i, \vec{y}_i\}_N) = \frac{1}{N} \sum_{i=1}^{N} \frac{1}{M} \sum_{j=1}^{M} \left| y_{ij} - f(\lambda_{ij},\vec{p}_i) \right|,
\end{equation}
and the Mean of Absolute Errors exceeding the 0.95 quantile (MAQE$_{0.95}$):
\begin{equation}
\text{MAQE}_{0.95}(\{(\vec{\lambda}, \vec{p})_i, \vec{y}_i\}_N) = \frac{1}{N_{0.95}} \sum_{k \in Q_{0.95}} \left| y_{k} - f(\lambda_{k}, \vec{p}_{k}) \right|,
\end{equation}
where $k \in Q_{0.95}$ denotes the indices corresponding to the top 5\% of the largest errors, and $N_{0.95}$ is the number of such indices. This latter metric aims to measure the mean of the largest errors, providing a more conservative assessment than the Maximum Absolute Error metric. A metric that focuses on the largest errors is informative, as in the case where the spectrum is dominated by the continuum, the MAE and MSE mostly measure the prediction in the continuum, not the lines. We expect errors in lines to be the largest errors, and at the same time the most correlated to uncertainties in abundance estimation. This is the reason why we expect MAQE$_{0.95}$ to be the metric most relevant for precise abundance inference.

As a final metric for a given model, we consistently report the smallest value obtained on the validation dataset during training. Thus, if the model begins to overfit, we do not use the overfitted model; instead, we select the model that achieved the lowest metrics on the validation dataset.

\subsection{Fine-tuning}

Fine-tuning is a strategy aimed at reducing the number of examples necessary to adapt a machine learning model to a new task or domain. In the context of stellar spectra emulation, we employ fine-tuning as a method to train effective emulators with significantly fewer synthetic spectra. The fine-tuning process involves a two-step strategy: initially, it is necessary to prepare a pre-trained model, which serves as a so-called base model. This model should be trained on data similar to the target domain, as this similarity is a fundamental condition for successful fine-tuning. Subsequently, data from the target domain are used to slightly modify the base model by continuing the training for some additional steps. Here, we used the simplest fine-tuning method, where all parameters of the model are updated, and all hyperparameters (e.g., maximum learning rate or training schedule) are kept unchanged from the pre-training phase.

Our base model underwent training for a million steps on an initial pre-training grid with fixed effective temperature and surface gravity. Subsequently, it was fine-tuned on a training grid where these parameters were variable. Because the training grid has two more free parameters, effective temperature and surface gravity, the models were minimally modified to allow different dimensionality of an input. It involved the modification of input layers of each model by appending two rows to input matrices. Although fine-tuning is usually run for fewer steps than pre-training, we decided to keep the number of training steps equal in both scenarios.

\subsection{Inference of parameters of stellar atmospheres}
\label{sec:inference_methods}

Various methods are used to infer the parameters of stellar atmospheres. Since the purpose of spectral parameter inference here was the validation of the emulators, simple optimization of the MSE between the predicted and true normalized spectra is a reasonable approach. As not all elemental abundances are constrained on the training grid, we first estimated the Cramér-Rao bounds to decide which parameters should be fitted and which should be excluded from inference. The purpose of this procedure was to restrict our results to elemental abundances for which inference can be expected to be more precise than 0.05 dex. This left us with 38 elemental abundances which were included in experiments of parameters fitting. It is not the purpose of this work to analyze these bounds in detail, so we do not discuss them further. For reference on how Cramér-Rao bounds might be applied in stellar spectroscopy, see \citet{2017ApJ...843...32T}.

We experimented with several optimizers to find the one that consistently fits the parameters of the stellar spectra from random initialization across the whole domain. The best was the Adam optimizer \citep{2014arXiv1412.6980K}, with gradient clipping set to 10.0 and a cosine rate scheduler with linear warm-up for the first 10\% of steps. For each spectrum, from a set of 256 spectra chosen randomly from the validation dataset, we repeated the optimization ten times with randomly chosen starting points for 2000 gradient updates and a maximum learning rate of 0.1. As a final result, we used the parameters with the smallest mean squared error fit.

\section{Results}
\label{sec:results}

\begin{figure*}
\resizebox{\hsize}{!}
{\includegraphics[clip]{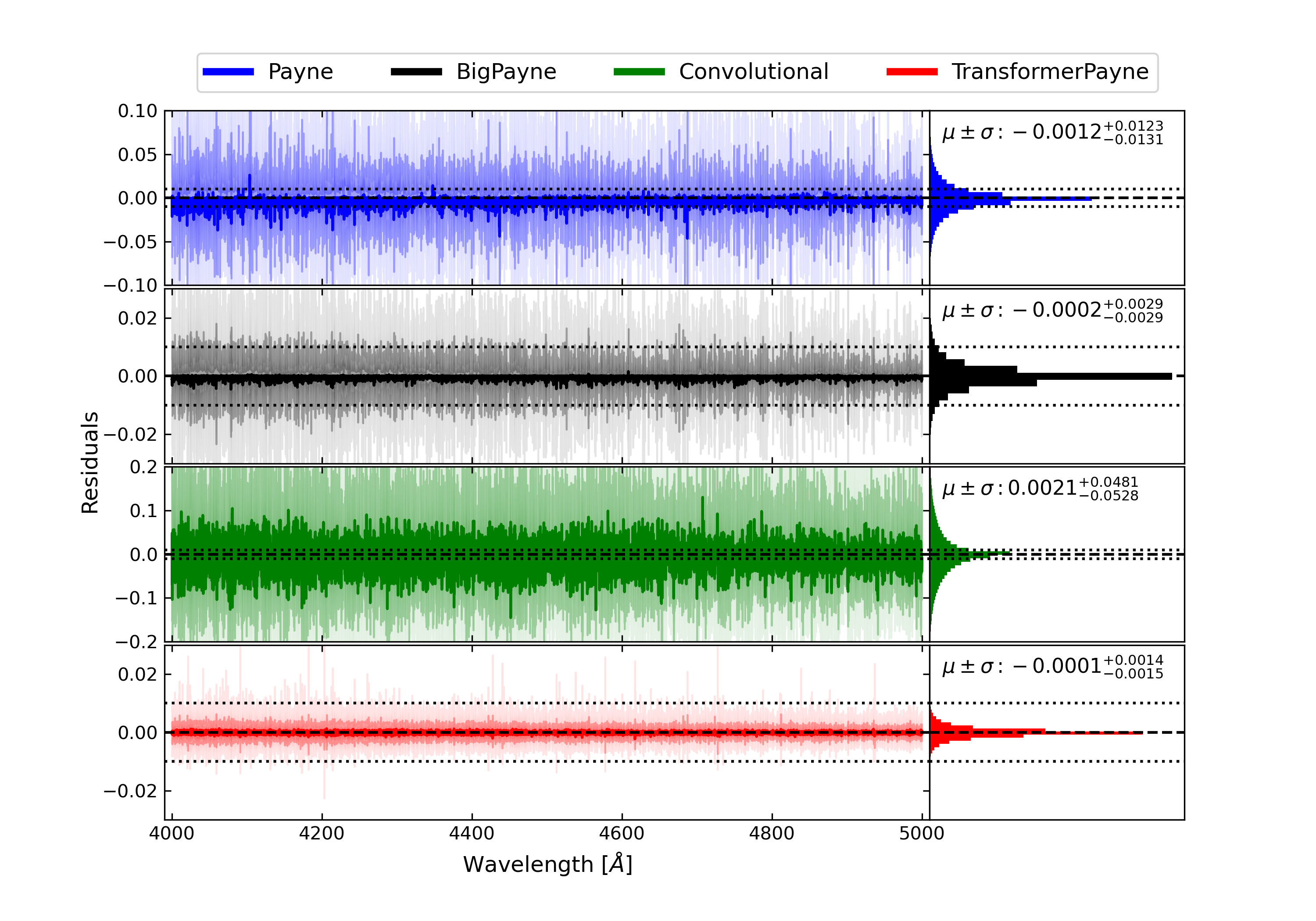}}
\caption{Residuals of the predictions of tested emulators on the validation dataset are depicted as a function of wavelengths (\textit{left}) and summarized in histograms (\textit{right}). In the left panels, the line indicates the median, while the bands correspond to 1$\sigma$ and 2$\sigma$ intervals. Specifically, the 1$\sigma$ interval is calculated using the 16th and 84th percentiles, and the 2$\sigma$ interval uses the 2.5th and 97.5th percentiles, representing dispersion around the median. Note that the panels in different rows do not share the same scale; however, dotted black lines indicate a 0.01 residual value in all panels. In the right panels, the summarizing statistics denoted as $\mu \pm \sigma$ report the median, along with the 16th and 84th percentiles. The TransformerPayne has the smallest bias and spread of residuals.}
\label{fig:fig4_residuals}
\end{figure*}

To demonstrate the strong inductive biases of the TransformerPayne, leading to better emulation, we comprehensively compared this model to our considered baselines in the following sections. We were also able to show how scaling improves the Payne emulator and the poor emulation quality of the Convolutional-based emulator.

In Sect.\,\ref{sec:training_from_scratch}, we present the results of training models from scratch on the training grid as a function of training set size and the number of training steps. Then, is Section\,\ref{sec:correlation_in_residuals} we investigated how residuals computed on spectra from validation dataset are correlated. Next, in Section\,\ref{sec:fine_tuning}, we describe the results of transfer learning for all models, using fine-tuning. Finally, in Section\,\ref{sec:fitting_spectral_parameters}, we present the results of the inverse problem of inferring parameters of stellar spectra using the best models.

\begin{figure*}
\resizebox{\hsize}{!}
{\includegraphics[clip]{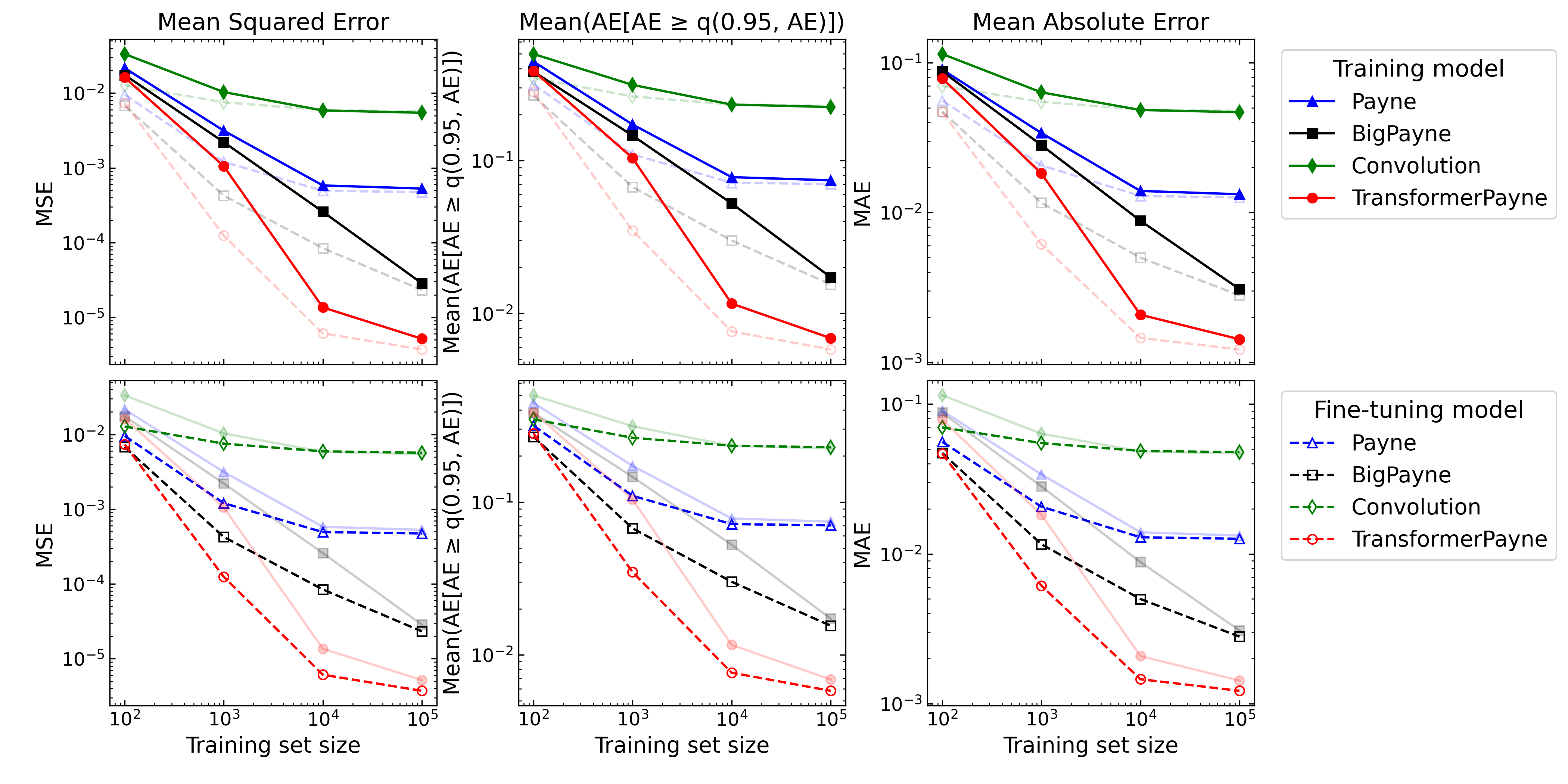}}
\caption{Results of training for four considered neural network emulators: The Payne, BigPayne, Convolution-based, and TransformerPayne. All panels display selected emulation metrics on a validation dataset as a function of training set size. Each training run was conducted for a fixed number of training steps equal one million. The metrics used are: Mean Squared Error (MSE), Mean of Absolute Errors above the 95th percentile of Absolute Error (MAQE$_{0.95}$), and Mean Absolute Error (MAE). The upper panels highlight the results of pre-training from scratch, while the lower panels highlight the results of fine-tuning.  Less expressive models, namely The Payne and Convolution-based models, saturate with a training set size of approximately $10^4$. In contrast, BigPayne and the TransformerPayne models exhibit no clear signs of saturation, even with the largest training set sizes. Except for the already saturated Payne and Convolution-based models, fine-tuned models surpass their counterparts trained from scratch. This improvement is most significant for training sets around 1000 spectra, where the enhancement ranges between 2 to 10 times, depending on the metric. The best emulator architecture is the TransformerPayne, which outperforms BigPayne by 2 to 10 times when the training set size ranges between $10^3$ and $10^5$. For the smallest training set size of $10^2$, all models perform comparably.}
\label{fig:fig5_legend_outsize}
\end{figure*}

\subsection{Training emulators from scratch on training grid}
\label{sec:training_from_scratch}

First, we present how neural networks compare to each other when trained from scratch on the training grid. Training from scratch involves optimizing the neural networks' parameters starting from their randomly initialized values. This contrasts with the fine-tuning approach, where the model is not initialized from random weights but from a state achieved after training on a pre-training grid. In this way, some features learned during pre-training can be reused in the fine-tuning process, improving the quality of the emulator and reducing the number of examples necessary to train the model. Experiments presented below shows the effect of scaling the size of the networks in the pair of The Payne and BigPayne, and the effect of different inductive biases when comparing the results of all emulators. 

\subsubsection*{Scaling training set size}

Training dataset size is known to have the biggest effect on the final precision of any machine learning model. Therefore, we started by measuring how the emulation accuracy changes with the training set size when the number of training steps is kept fixed and equal one million. We scaled the training set size by sampling from 100 to 100\,000 spectra from the training dataset. Models that scale well with data size typically have appropriate inductive biases and usually perform well in transfer learning scenarios.

All models consistently improved their emulation quality, as measured by the metrics used, though the rate of improvement varied. For the smallest training dataset, all methods predicted spectra comparably well, with a Mean Absolute Error on the order of 0.1. However, when scaling up the training set size, the emulation quality of Convolutional-based and The Payne models increased but showed signs of saturation. By saturation, we refer to the plateau in improvement between training set sizes of 10\,000 and 100\,000, where increasing the training set size had diminishing returns. In contrast, for the BigPayne and TransformerPayne models, there were no signs of saturation up to the use of the full training set with 100\,000 spectra. The TransformerPayne model improved at a slightly faster rate, as measured by these metrics, outperforming all other emulators by a factor of between 3 and 10, depending on the metric used. The best model shows an MSE around $4\times10^{-6}$, MAE around $10^{-3}$, and MAQE$_{0.95}$ around $6\times 10^{-3}$. All these results are presented in the upper panels of Fig.\,\ref{fig:fig5_legend_outsize}.

The residuals obtained from the predictions of the best emulators trained from scratch for the spectra from a validation dataset are depicted in Fig.\,\ref{fig:fig4_residuals}. In the left panels, this figure shows residuals as a function of wavelength, summarized using the median, and 1$\sigma$ and 2$\sigma$ bands (estimated using the 16th and 84th percentiles for the former, and the 2.5th and 97.5th percentiles for the latter). In the right panel, there are histograms summarizing the distribution of residuals globally, together with the statistics $\mu \pm \sigma$ showing the median, along with the 16th and 84th percentiles. Mean residuals and spread align with the same model ranking, with the best being TransformerPayne ($-0.0001^{+0.0014}_{-0.0015}$), followed by BigPayne ($-0.0002^{+0.0029}_{-0.0029}$), The Payne ($-0.0012^{+0.0123}_{-0.0131}$) and Convolutional-based emulator ($0.0021^{+0.0481}_{-0.0528}$). The Payne results concur with the observation that it tends to saturate, with a Mean Absolute Error of emulation of normalized flux on the order of 0.01.

Finally, an overview of the emulation quality of the two best models, TransformerPayne and BigPayne, can be seen in Fig.\,\ref{fig:fig1_residuals_4100_4130}. It shows the emulation in the narrow wavelength range from 4100\,\r{A} to 4130\,\r{A} of an example spectrum from the validation dataset. It illustrates that TransformerPayne has much smaller residuals, with the differences being most prominent in narrow spectral lines.

\subsubsection*{Scaling number of training steps}

\begin{figure*}
\resizebox{\hsize}{!}
{\includegraphics[clip]{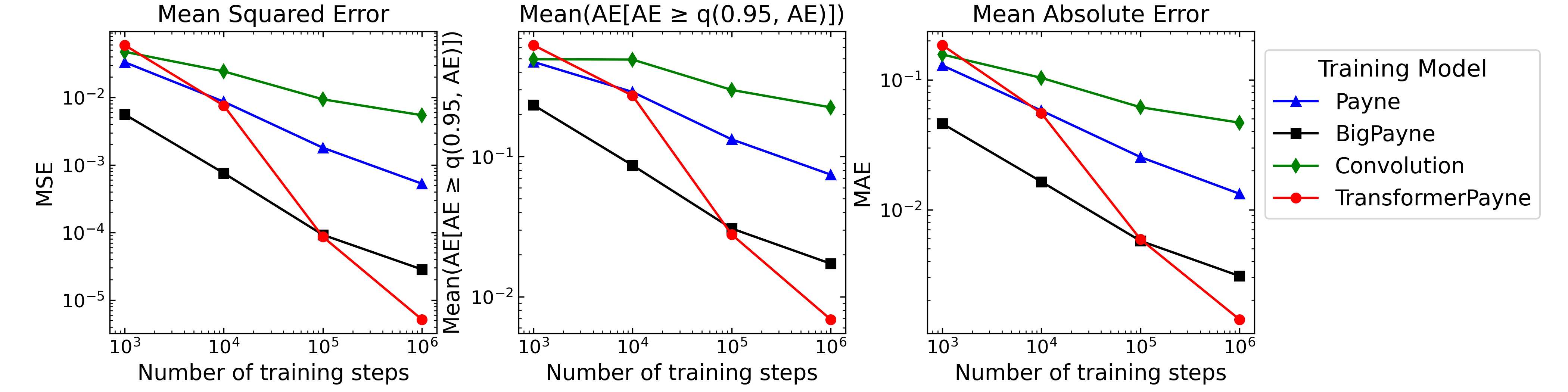}}
\caption{Emulation metrics as a function of number of training steps with a full training dataset containing 100\,000 stellar spectra. The considered emulators continue to decrease prediction errors with an increased number of training steps. The ranking of the Convolutional-based, BigPayne, and the Payne emulators remains unchanged as training steps increase, with their accuracy improving three-fold, ten-fold, and fifteen-fold respectively, as measured using MAE. TransformerPayne exhibits qualitatively different behavior, performing the worst at 1,000 steps but improving by 130 times when trained for a million steps, resulting in a best MAE of approximately 0.0015. This shows that, with sufficient computational resources, TransformerPayne outperforms other models in terms of spectral emulation.}
\label{fig:fig6_scalling_steps}
\end{figure*}

Length of the training is another important aspect of scaling laws for large neural networks. If the improvement from longer training is not plateauing, then with smaller, more efficient model trained for more steps, we can obtain the same accuracy as for larger models \citep{2022arXiv220315556H}. This motivates the experiments with scaling the number of training steps. The training step is a single update of free parameters of a model. The corrections are calculated based on a batch of data, which is this work contains 16 spectra. We decided to run training of models spanning three orders of magnitude in the number of training steps, from a thousand to a million steps, while keeping the number of stellar spectra in the training dataset fixed at 100\,000.

First, all models continued to improve without showing any signs of saturation when trained for up to 1 million steps. We can differentiate between three groups of models. The Convolutional-based model improves its emulation quality with longer training, from about 0.15 to 0.05 in MAE, when increasing the number of training steps from 1000 to 1\,000\,000. This is the smallest relative improvement among all considered models. The Payne and BigPayne models appear to improve at a similar rate. The Payne model reduces MAE approximately tenfold, from 0.1 to 0.01, while BigPayne reduces MAE from 0.045 to 0.003, which is a fifteenfold improvement. The larger model consistently outperforms the smaller one by approximately three times.

In contrast, the TransformerPayne model exhibits the biggest relative improvement, from a MAE of 0.2 when trained for a thousand steps, to 0.0015, which is approximately 130 times smaller than the initial emulation error. When trained for thousand steps, its metrics are the worst among all the models. When trained for hundred thousand steps, it matches the performance of BigPayne and surpasses it by two times at million training steps. It is worth highlighting that the scaling of this model is not saturating. If this trend holds when scaling training to 10 million steps, the difference with respect to BigPayne is expected grow further.

Mean Squared Error and the Mean of Absolute Errors above the 95th percentile of Absolute Error show qualitatively the same picture of scaling with the length of training. TransformerPayne is the only model that shows an accuracy better than 0.01 when measuring MAQE$_{0.95}$, which is the most sensitive to errors in spectral lines. It is also worth noting that while TransformerPayne is slightly worse regarding MAE when training for 100\,000 steps, it is slightly better in terms of MAQE$_{0.95}$. This means that the advantage of TransformerPayne is most prominent in spectral lines. This experiment is summarized in detail in Fig.\,\ref{fig:fig6_scalling_steps}.

\subsection{Correlations in the residuals of emulators' predictions}
\label{sec:correlation_in_residuals}

When considering inductive bias for stellar spectra, it should be tailored to efficiently handle long-range correlations between spectral features, such as lines of the same ion. This means that a model with good inductive bias should internally learn features optimized for many related spectral lines, even if they are widely separated. This will lead to more accurate predictions of flux in those lines but also to more correlated errors. This can be observed by analyzing the correlation in residuals obtained as the difference between emulators' predictions and spectra from the validation dataset.

\begin{figure*}
\centering
\resizebox{0.95\hsize}{!}
{\includegraphics[clip]{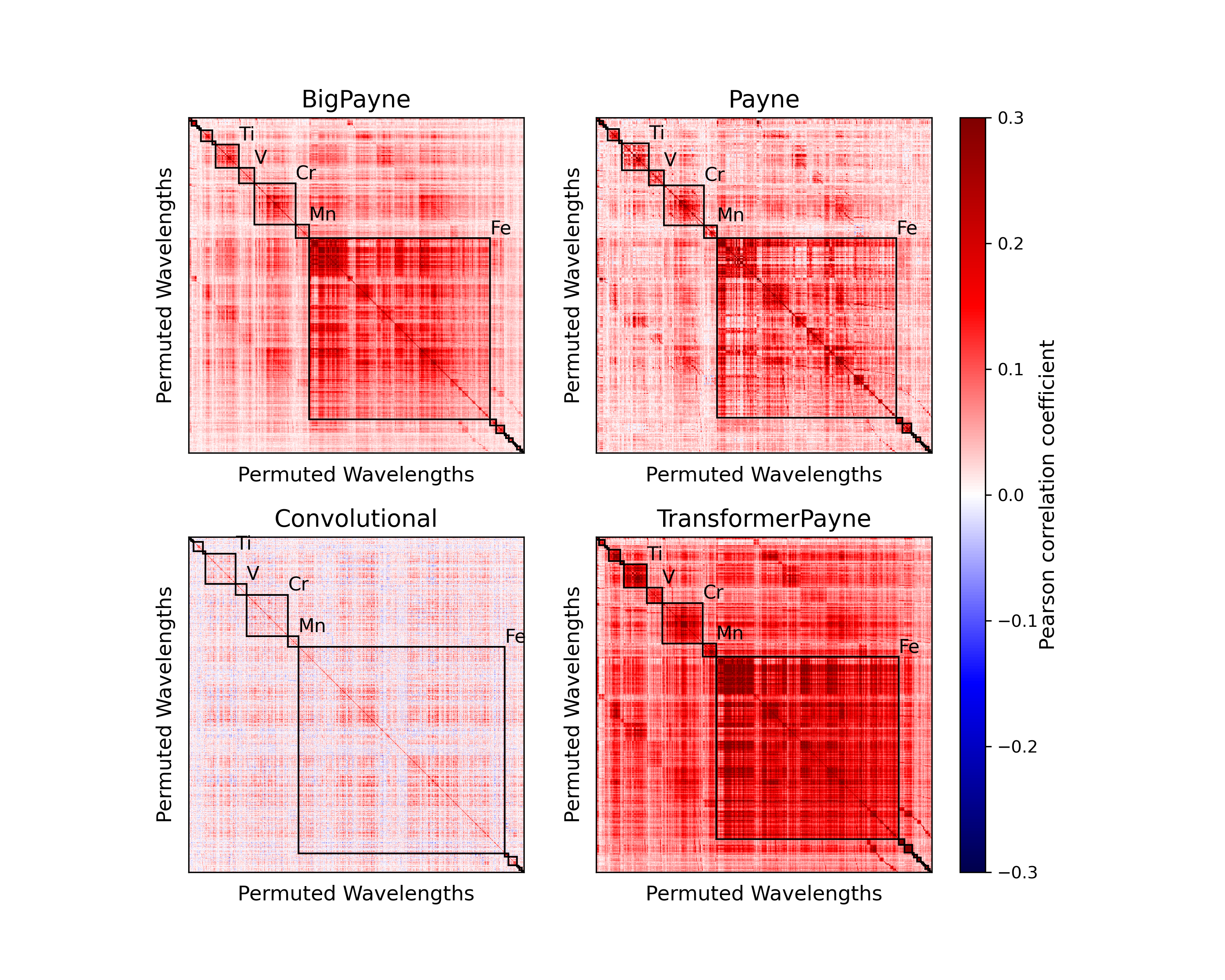}}
\caption{Pearson correlation over residuals sorted by minimal mean gradient computed over ten spectra from validation dataset. The squares in black show the parts corresponding to wavelengths dominated by a single element, e.g. Fe dominates the most of the spectrum, and Cr is the second most represented element in considered spectral range. In each square, the residuals are further sorted by the second most relevant gradients, and so forth, so that we also visualize the influence due to blended features. The structures in every square illustrate the influence of blends on the residuals' correlations. The median of correlation matrix is the largest for the TransformerPayne emulator and equals approximately 0.094, while equal 0.046 for the BigPayne. The stronger correlation shows that TransformerPayne harnesses the long range information from the spectra for better spectral emulation.}
\label{fig:fig7_corr_matrix_residuals}
\end{figure*}

To measure the correlation we calculated sample Pearson correlation matrix, $r_{xy}$, of the vectors of residuals from all 1024 validation samples for every pair of wavelengths. Sample Pearson correlation matrix for all emulators are shown in Fig.\,\ref{fig:fig7_corr_matrix_residuals}. The median of the correlation matrix is approximately 0.094 for TransformerPayne, 0.046 for BigPayne, 0.038 for The Payne, and 0.005 for the Convolutional-based emulator. The correlation in residuals increases as the emulator becomes more accurate and is the largest for TransformerPayne. The correlations in residuals are 1.2 times stronger in BigPayne compared to The Payne, and 2 times stronger in TransformerPayne compared to BigPayne. Emulation quality improves by a factor of 3.5 when moving from The Payne to BigPayne, and by a factor of 2 when moving from BigPayne to TransformerPayne. This indicates that increase in correlation is primarily due to the distinct inductive biases inherent in TransformerPayne, rather than a result of better emulation accuracy.

To check if correlations are associated with spectral features corresponding to a fixed element, we assigned the abundance to each wavelength and then permuted the rows and columns of the correlation matrix using these labels. At each wavelength, we associate the pixel to the abundance based on the minimal gradient (as when abundance increases, the normalized flux in spectral lines usually decreases) computed across ten random spectra from the validation dataset. This aims to assign to every wavelength the elemental abundance that mostly affects the flux. For better visualization we put boxes on blocks of permuted correlation matrix with the same primary label, see Fig.\,\ref{fig:fig7_corr_matrix_residuals}. For example iron is dominating the most of the wavelengths and chromium is the second most prominent element. As can be seen, the correlations have a clear structure associated with the elemental abundances that primarily influence the normalized flux at a given wavelength. The secondary structure within each box may result from blends with other lines.

Notably, the small dark boxes represent elements with few lines in the spectra, indicating the highest correlation for these elements. As shown in Fig.\,\ref{fig:fig7_corr_matrix_residuals}, this effect is most pronounced for the TransformerPayne emulator. This further supports the claim that TransformerPayne more effectively leverages the long-range correlations present in stellar spectra.

\subsection{Generalizing from pre-training dataset}
\label{sec:fine_tuning}

In a fine-tuning experiment, we test how well the emulator can adapt to different spectral types or, more generally, different domains when pre-trained on a smaller grid or a grid with simplified physics. TransformerPayne, which has inductive biases particularly well suited for stellar spectra emulation, should generalize relatively well, but this strategy can be used for all considered emulators.

Pre-training was run for one million training steps on the pre-training grid, which differs from the training grid by having the effective temperature and surface gravity fixed. Then we fine-tuned base models over one million steps using subsets of the training grid ranging in size from 100 to 100\,000 spectra. In all cases, the fine-tuning approach yielded models that were either better than or comparable to those trained from scratch. 

Results close to the baseline were observed for models showing signs of saturation, which are The Payne and the Convolution-based model, when fine-tuning with dataset sizes of 10\,000 and 100\,000. Similarly, the benefit of the pre-training strategy decreased as the size of the fine-tuning dataset approached that of the pre-training dataset. When fine-tuning the models on 100 spectra, the improvement from fine-tuning is modest, and improve the emulation as measured using mean absolute error a slightly less than two times, e.g., for BigPayne it reduces the MAE from 0.09 to 0.05, and comparably for other emulators.

The fine-tuning shows best results when applied to datasets ranging in size from 1000 to 10\,000 examples. For a fixed training size, the emulation quality of BigPayne and TransformerPayne models improves by a factor from 1.5 to 10, depending on the metric. As a representative example, for mean absolute error and TransformerPayne the improvement is from 0.02 to 0.006 when fine-tuning on 1000 spectra, and from 0.002 to 0.0015 for fine-tuning grid with 10\,000 spectra. The BigPayne also show comparable relative improvements, from 0.03 to 0.01 for 1000 spectra and from 0.009 to 0.005 for 10\,000 spectra. 

When considering the target emulation metric fixed, for instance, MAE equal to 0.005, training BigPayne from scratch requires about 40\,000 synthetic spectra. When using fine-tuning, having four times fewer spectra in the targeted grid gives the same results. To train TransformerPayne to this accuracy, we need approximately 4000 spectra, and with fine-tuning, only 2000 spectra. Together, the application of the TransformerPayne architecture with fine-tuning enables training emulators of the same emulation accuracy with 20 times less data. The details of these results, as measured using MAE and other considered metrics, are illustrated in the bottom panels of Fig.\,\ref{fig:fig5_legend_outsize}. 

\subsection{Inference of stellar spectra parameters using emulators}
\label{sec:fitting_spectral_parameters}

A key application of emulators of stellar spectra is to amortize the cost of calculating synthetic spectra using spectral synthesis codes in the inference pipelines of large astronomical surveys. The goal is to consistently and accurately infer parameters of stellar atmospheres, and a proper analysis pipeline requires the systematic due to error in emulation to be subdominant compared to either the statistical error from the photon noise or other systematics. For example, when we include non-LTE physics in the synthetic spectrum modeling, this might affect inferred abundances on the order of 0.1\,dex. In such cases, we expect the systematic due to the emulator to be negligible compared to 0.1\,dex.

Until now, we were reporting metrics that are easily measurable during the training of emulators but are not directly informative regarding the accuracy and precision of the parameters we aim to infer. To address this we fitted effective temperature, surface gravity and 38 individual abundances of stellar atmospheres for 256 spectra from validation dataset. We fitted 38 individual abundances, not all 98, as only 38 can be accurately inferred given the parameters and wavelength range covered by training grid (for details see Sect.\,\ref{sec:inference_methods}). Inference used optimization of the mean squared error of prediction of emulators, and we report parameters with smallest MSE as a derived estimate. The result of this fitting is summarized using standard deviation of errors in Tables\,\ref{tab:fit_results_1} and \ref{tab:fit_results_2}.

Table\,\ref{tab:fit_results_1} shows summary calculated over all spectra for effective temperature, surface gravity and helium abundance. The TransformerPayne results are the best for all these parameters, $T_{\text{eff}}$ ($\sigma = 3.70$ [K]), $\log g$ ($\sigma = 0.005$) and $N_{\text{He}} / N_{\text{tot}}$ ($\sigma = 0.003$). The second best is BigPayne, followed by Convolutional-based emulator and The Payne.

The results for individual abundances for all other elements, $\text{[X/H]}$, are summarized in Table \ref{tab:fit_results_2}. Here, the standard deviation of the inferred abundances is calculated in three different abundance groups, as the inference accuracy can vary substantially at different abundance levels. Nonetheless, regardless of the abundance group, TransformerPayne, with just two exceptions, is the most precise emulator. One representative example is iron abundance [Fe/H]. In all three abundance levels: $-2.0 \le \text{[Fe/H]} < -1.0$, $-1.0 \le \text{[Fe/H]} < 0.0$, and $0.0 \le \text{[Fe/H]} < 1.0$, the most precise abundances are obtained using the TransformerPayne emulator, with precisions equal to 0.0042, 0.0038, and 0.0045, respectively. The other emulators, in order of decreasing precision, are BigPayne (0.0060, 0.0050, and 0.0073, respectively), Convolutional-based (0.0192, 0.0175, and 0.0204, respectively), and The Payne (0.0516, 0.0338, and 0.0333, respectively). As iron lines are prominent even at the lowest level of iron abundance, the precision of inferred abundance is comparable across these three levels. When investigating the case of elements that have only a few weak lines in the spectrum, like aluminum, the precision of inferred abundances can greatly vary depending on the true abundance level, being high when the true abundance is high but degrading in the case of low abundance. Analogously to iron, in the three abundance levels: $-2.0 \le \text{[Al/H]} < -1.0$, $-1.0 \le \text{[Al/H]} < 0.0$, and $0.0 \le \text{[Al/H]} < 1.0$, the precision of inferred $\text{[Al/H]}$ is best for TransformerPayne, with values equal to 0.3713, 0.1884, and 0.0161, respectively. Other emulators, also in order of decreasing precision, are BigPayne (0.5278, 0.4038, and 0.0332, respectively), Convolutional-based (0.7450, 0.7979, and 0.7952, respectively), and The Payne (0.9599, 0.9245, and 0.7982, respectively). The mean relative improvement of abundances accuracy from TransformerPayne emulator compared to other models is reported in Table \ref{tab:relative_improvement_abundances}.

\begin{figure*}
\centering
\resizebox{0.95\hsize}{!}
{\includegraphics[clip]{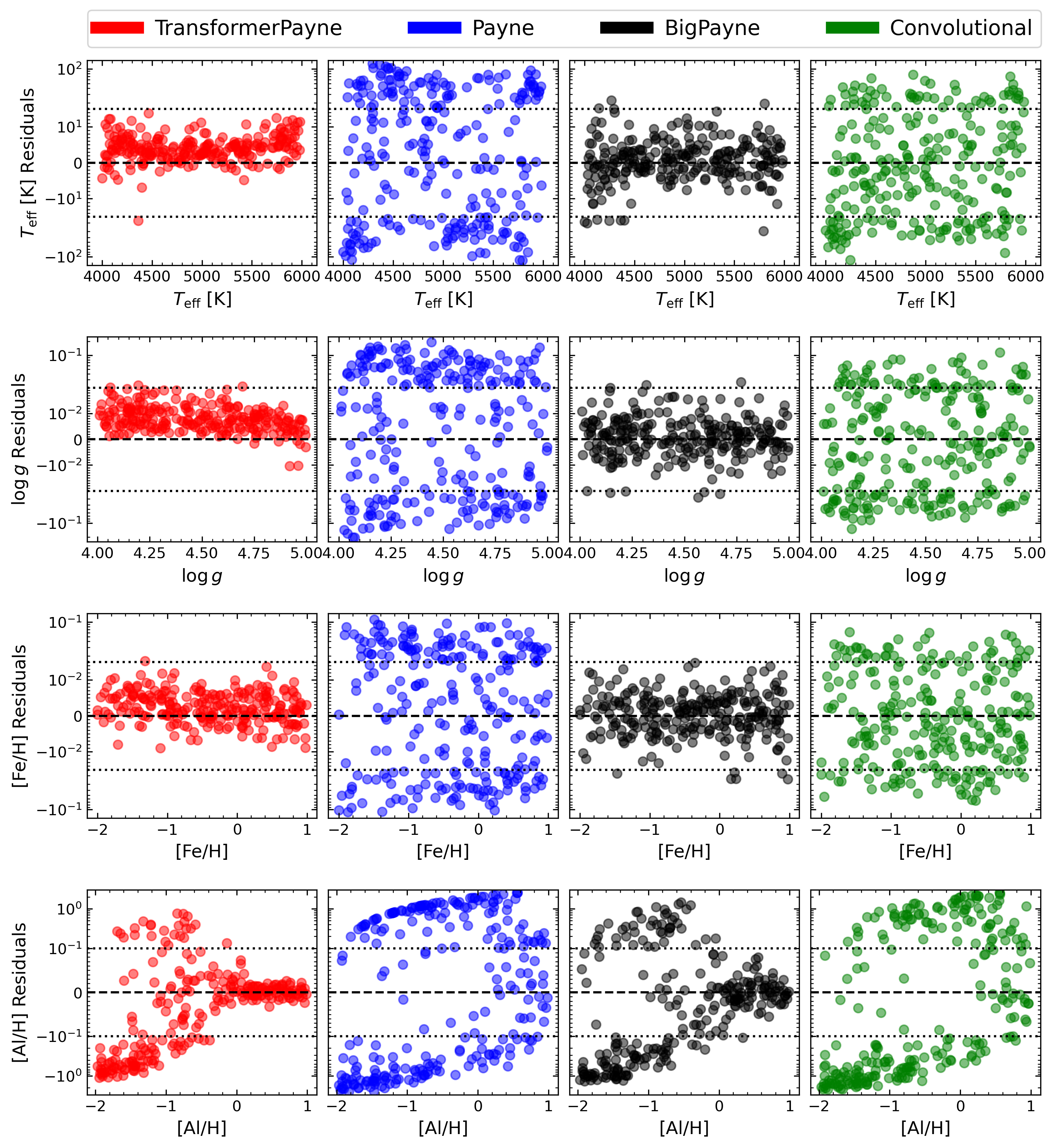}}
\caption{The recovery of effective temperature, $T_{\text{eff}}$, logarithm of surface gravity, $\log g$, iron abundance, $\text{[Fe/H]}$, and aluminum abundance, $\text{[Al/H]}$ of the validation set through fitting (from \textit{top} to \textit{bottom} row, respectively) with all considered emulators. Note that the scale is linear between the dotted lines and logarithmic outside of them, and that the scale is shared for panels in each row. TransformerPayne offers the most precise predictions; however, BigPayne tends to exhibit smaller biases, though both are negligible. This is most prominent for $\log g$, where TransformerPayne bias equals 0.008 and BigPayne bias equals 0.001. Particularly interesting is the result for [Al/H], where for lower true abundances, the information on Aluminum abundance subsides, which prominently affects the precision of TransfomerPayne and BigPayne.
}
\label{fig:fig8_residuals_some_params}
\end{figure*}

Illustrative results of fitting the effective temperature, logarithm of surface gravity, and the abundance of iron and aluminum are presented in Fig.\,\ref{fig:fig8_residuals_some_params}. As confirmed in Table\,\ref{tab:fit_results_2}, TransformerPayne is the most precise emulator for all these parameters. Figure\,\ref{fig:fig8_residuals_some_params} shows that, generally, TransformerPayne has larger biases than BigPayne. The presence of bias is related to the correlated errors discussed in Sect.,\ref{sec:correlation_in_residuals} and forms a trade-off: efficient handling of long-range dependence between lines that characterized TransformerPayne leads to more precise emulation, but at the same time, it causes correlated errors which lead to correlated biases in inferred abundances. It is worth noting that these biases are still at least an order of magnitude smaller than the biases due to the approximations usually used in synthetic spectra calculations and can be easily corrected for. Finally, the last row depicts the residuals of [Al/H] abundance, showing the dependence of accuracy on the true abundance. When the lines of Al become very weak in the spectra, comparable to the emulator precision, then the inferred abundance is no longer precise. Next to aluminum, this is also the case for many elements, such as Na, K, Zr, or Nb.

\begin{table*}

\caption{Summary of the effective temperature, logarithm of surface gravity, and helium abundance recovery. Shown are the standard deviations of the recovery. The results are shown for each parameter. Transformer Payne is denoted as TP, the Payne as P, BigPayne as BP, and the Convolutional-based emulator as C. The results with the smallest standard deviation for each parameter are given in bold.}             

\label{tab:fit_results_1}
\begin{center}
{\small
\begin{tabular}{l|rrrr}
\hline
Parameter & TransformerPayne (TP) & The Payne (P) & BigPayne (BP) & Convolutional (C) \\
\hline
     $T_\textrm{eff}$ [K] & \textbf{3.7027} &    43.3008 &          6.3656 &    \multicolumn{1}{r}{24.3971}  \\
     $\log g$ &          \textbf{0.0050} &     0.0632 &          0.0080 &     \multicolumn{1}{r}{0.0330}  \\
    $N_{\text{He}} / N_{\text{tot}}$ & \textbf{0.0029} &     0.0213 &          0.0034 &     \multicolumn{1}{r}{0.0136} \\
\hline
\end{tabular}
}
\end{center}

\vspace{0.5cm}

\begin{center}
\caption{Summary of the individual abundance recoveries, including the standard deviations. The results are shown for three different metallicity ranges. Transformer Payne is abbreviated as TP, the Payne as P, BigPayne as BP, and the Convolutional-based emulator as C. For each abundance range, the results with the smallest standard deviation are in bold.}
\label{tab:fit_results_2}
{\small
\begin{tabular}{l|rrrr|rrrr|rrrr}
\hline
  & \multicolumn{4}{c|}{$-2.0 \le \text{[X/H]} < -1.0$} & \multicolumn{4}{c|}{$-1.0 \le \text{[X/H]} < 0.0$} & \multicolumn{4}{c}{$0.0 \le \text{[X/H]} \le 1.0$} \\
\multicolumn{1}{l|}{[X/H]} & TP & P & BP & C & TP & P & BP & C & TP & P & BP & C \\
\hline
        C & \textbf{0.4933} &    0.6964 &          0.5955 &    0.7195 & \textbf{0.3920} &       0.6510 &        0.4557 &       0.6436 & \textbf{0.4629} &     0.7459 &          0.5566 &     0.8036 \\
       Na & \textbf{0.0313} &    0.2824 &          0.0493 &    0.1618 & \textbf{0.0076} &       0.1678 &        0.0167 &       0.1484 & \textbf{0.0070} &     0.0592 &          0.0102 &     0.0324 \\
       Mg & \textbf{0.0058} &    0.1444 &          0.0142 &    0.0689 & \textbf{0.0052} &       0.0759 &        0.0109 &       0.0300 & \textbf{0.0068} &     0.0599 &          0.0081 &     0.0317 \\
       Al & \textbf{0.3713} &    0.9599 &          0.5278 &    0.7450 & \textbf{0.1884} &       0.9245 &        0.4038 &       0.7979 & \textbf{0.0161} &     0.7982 &          0.0332 &     0.7952 \\
       Si & \textbf{0.0096} &    0.2231 &          0.0244 &    0.1295 & \textbf{0.0089} &       0.1482 &        0.0223 &       0.0922 & \textbf{0.0077} &     0.0996 &          0.0133 &     0.0624 \\
        K & \textbf{0.2222} &    0.5804 &          0.2996 &    0.5176 & \textbf{0.0208} &       0.3401 &        0.0423 &       0.3774 & \textbf{0.0230} &     0.4331 &          0.0415 &     0.5008 \\
       Ca & \textbf{0.0057} &    0.0829 &          0.0092 &    0.0306 & \textbf{0.0054} &       0.0498 &        0.0077 &       0.0204 & \textbf{0.0052} &     0.0406 &          0.0064 &     0.0210 \\
       Sc & \textbf{0.0105} &    0.1320 &          0.0183 &    0.0549 & \textbf{0.0090} &       0.0577 &        0.0103 &       0.0368 &          0.0090 &     0.0504 & \textbf{0.0088} &     0.0301 \\
       Ti & \textbf{0.0052} &    0.0579 &          0.0073 &    0.0231 & \textbf{0.0050} &       0.0466 &        0.0062 &       0.0236 & \textbf{0.0053} &     0.0331 &          0.0055 &     0.0176 \\
        V & \textbf{0.0064} &    0.1347 &          0.0275 &    0.0766 & \textbf{0.0055} &       0.0529 &        0.0069 &       0.0260 & \textbf{0.0055} &     0.0359 &          0.0071 &     0.0231 \\
       Cr & \textbf{0.0057} &    0.0734 &          0.0098 &    0.0355 & \textbf{0.0045} &       0.0425 &        0.0068 &       0.0176 & \textbf{0.0051} &     0.0337 &          0.0057 &     0.0181 \\
       Mn & \textbf{0.0068} &    0.1227 &          0.0142 &    0.0319 & \textbf{0.0053} &       0.0445 &        0.0072 &       0.0211 & \textbf{0.0059} &     0.0390 &          0.0074 &     0.0191 \\
       Fe & \textbf{0.0042} &    0.0516 &          0.0060 &    0.0192 & \textbf{0.0038} &       0.0338 &        0.0050 &       0.0175 & \textbf{0.0045} &     0.0333 &          0.0073 &     0.0204 \\
       Co & \textbf{0.0070} &    0.1172 &          0.0127 &    0.0444 & \textbf{0.0064} &       0.0594 &        0.0106 &       0.0283 & \textbf{0.0068} &     0.0505 &          0.0074 &     0.0223 \\
       Ni & \textbf{0.0065} &    0.1316 &          0.0203 &    0.0569 & \textbf{0.0059} &       0.0528 &        0.0090 &       0.0215 & \textbf{0.0064} &     0.0421 &          0.0080 &     0.0208 \\
       Cu & \textbf{0.2972} &    0.6359 &          0.3958 &    0.7370 & \textbf{0.1224} &       0.3658 &        0.1314 &       0.6836 & \textbf{0.0182} &     0.2388 &          0.0358 &     0.6395 \\
       Zn & \textbf{0.0714} &    0.2948 &          0.1088 &    0.2148 & \textbf{0.0176} &       0.1056 &        0.0343 &       0.1077 & \textbf{0.0177} &     0.1016 &          0.0274 &     0.0594 \\
       Ga & \textbf{0.1517} &    0.3648 &          0.2169 &    0.4818 & \textbf{0.0435} &       0.4284 &        0.1056 &       0.4384 & \textbf{0.0300} &     0.3733 &          0.0642 &     0.5176 \\
       Sr & \textbf{0.0112} &    0.1731 &          0.0350 &    0.1047 & \textbf{0.0105} &       0.0929 &        0.0154 &       0.0434 & \textbf{0.0085} &     0.0643 &          0.0119 &     0.0378 \\
        Y & \textbf{0.0104} &    0.1236 &          0.0320 &    0.1024 & \textbf{0.0088} &       0.0658 &        0.0109 &       0.0395 & \textbf{0.0104} &     0.0594 &          0.0120 &     0.0369 \\
       Zr & \textbf{0.0233} &    0.2264 &          0.0687 &    0.1736 & \textbf{0.0071} &       0.0878 &        0.0087 &       0.0347 & \textbf{0.0065} &     0.0481 &          0.0072 &     0.0254 \\
       Nb & \textbf{0.3169} &    0.7815 &          0.3434 &    0.4919 & \textbf{0.1655} &       0.5468 &        0.1822 &       0.3982 & \textbf{0.0130} &     0.6823 &          0.0259 &     0.4520 \\
       Ru & \textbf{0.2455} &    0.5599 &          0.3095 &    0.4894 & \textbf{0.0474} &       0.3482 &        0.1879 &       0.4093 & \textbf{0.0131} &     0.1457 &          0.0174 &     0.3562 \\
       Ba & \textbf{0.0135} &    0.1959 &          0.0250 &    0.0701 & \textbf{0.0128} &       0.0973 &        0.0184 &       0.0579 & \textbf{0.0145} &     0.0903 &          0.0170 &     0.0527 \\
       La & \textbf{0.0500} &    0.2509 &          0.1310 &    0.2539 & \textbf{0.0103} &       0.1054 &        0.0174 &       0.1024 & \textbf{0.0111} &     0.0536 &          0.0114 &     0.0441 \\
       Ce & \textbf{0.0660} &    0.4129 &          0.1657 &    0.3650 & \textbf{0.0086} &       0.2088 &        0.0241 &       0.1133 & \textbf{0.0070} &     0.0462 &          0.0078 &     0.0372 \\
       Pr & \textbf{0.1723} &    0.6660 &          0.2569 &    0.5028 & \textbf{0.0152} &       0.5631 &        0.0674 &       0.3206 & \textbf{0.0110} &     0.2041 &          0.0140 &     0.0976 \\
       Nd & \textbf{0.0291} &    0.2792 &          0.1041 &    0.2328 & \textbf{0.0078} &       0.1715 &        0.0139 &       0.0721 & \textbf{0.0088} &     0.0534 &          0.0094 &     0.0360 \\
       Sm & \textbf{0.1946} &    0.4752 &          0.2386 &    0.4656 & \textbf{0.0310} &       0.3475 &        0.0797 &       0.5090 & \textbf{0.0121} &     0.0999 &          0.0147 &     0.3127 \\
       Eu & \textbf{0.0722} &    0.3852 &          0.1628 &    0.3429 & \textbf{0.0170} &       0.2064 &        0.0287 &       0.2656 & \textbf{0.0121} &     0.1471 &          0.0153 &     0.0534 \\
       Gd & \textbf{0.2229} &    0.5599 &          0.2916 &    0.5211 & \textbf{0.0218} &       0.3666 &        0.0693 &       0.4981 & \textbf{0.0145} &     0.1181 &          0.0170 &     0.2020 \\
       Dy & \textbf{0.1683} &    0.4222 &          0.2360 &    0.4283 & \textbf{0.0166} &       0.1962 &        0.0455 &       0.3086 & \textbf{0.0130} &     0.1654 &          0.0161 &     0.1171 \\
       Ho & \textbf{0.4148} &    0.5983 &          0.4638 &    0.8137 & \textbf{0.3495} &       0.4725 &        0.3968 &       0.8644 & \textbf{0.0259} &     0.3495 &          0.0346 &     0.8525 \\
       Er & \textbf{0.2974} &    0.4970 &          0.3836 &    0.6850 & \textbf{0.0275} &       0.3015 &        0.1746 &       0.5148 & \textbf{0.0180} &     0.2149 &          0.0214 &     0.5654 \\
        W & \textbf{0.3254} &    0.5269 &          0.4879 &    0.7095 & \textbf{0.2087} &       0.4003 &        0.2488 &       0.8310 & \textbf{0.0221} &     0.3969 &          0.0347 &     0.7401 \\
       Os & \textbf{0.3399} &    0.5532 &          0.4703 &    0.6630 & \textbf{0.1444} &       0.4306 &        0.2994 &       0.6425 & \textbf{0.0222} &     0.4225 &          0.0389 &     0.3821 \\
       Pb &          0.2567 &    0.4819 & \textbf{0.2499} &    0.6262 & \textbf{0.1325} &       0.4000 &        0.1475 &       0.6235 & \textbf{0.0250} &     0.3252 &          0.0348 &     0.5954 \\
\hline
\end{tabular}
}
\end{center}
\end{table*}

\begin{table}

\caption{Summary of the improvement of average abundance precision of TransformerPayne with respect to the Payne, BigPayne, and Convolutional-based emulators across three metallicity ranges. This quantifies how much smaller the systematic errors are in the final estimation due to imperfect emulation when using TransformerPayne compared to other emulators.}
\label{tab:relative_improvement_abundances}
\begin{center}
{ \footnotesize
\begin{tabular}{rrrrr}
\hline
Abundance range & The Payne & BigPayne & Convolutional \\
\hline
$-2.0 \le \text{[X/H]} < -1.0$    & $\times~8.49$ & $\times~1.89$ & $\times~4.93$ \\
$-1.0 \le \text{[X/H]} < -0.0$ & $\times~10.47$ & $\times~1.98$ & $\times~8.37$ \\
$0.0 \le \text{[X/H]} \le 1.0$   & $\times~11.77$ & $\times~1.37$ & $\times~12.72$ \\
\hline
\end{tabular}
}
\end{center}
\end{table}

\section{Discussion}
\label{sec:discussion}

In this study, we develop the TransformerPayne architecture, which demonstrates superior emulation quality in modeling complex stellar spectra. TransformerPayne reduces Mean Absolute Errors of emulation tenfold when compared to the Payne emulator, and twofold relative to BigPayne. And at a fixed emulation precision, TransformerPayne enables training with ten times fewer spectra compared to BigPayne. TransformerPayne also demonstrates improved performance in generalizing from fixed temperature and surface gravity to the grid where those parameters vary. This generalization makes it possible to use a fine-tuning approach and train the emulator with at least two times fewer spectra. The better emulation also leads to better recovery of the labels. On average, TransformerPayne's precision for inferred abundances is between 8.49 and 11.77 times better than The Payne and between 1.37 and 1.98 times better than BigPayne. This proves that both scale and appropriate architecture is important for precise emulation of stellar spectra.

\subsection{TransformerPayne -- appropriate inductive biases and scaling}

The superiority of TransformerPayne in emulating spectra can be attributed to its appropriate inductive biases. The explicit parametrization in wavelengths enables the model to learn features shared across lines of the same element, even if they are widely separated in wavelength. Additionally, the attention mechanism captures this long-range information by conditioning the wavelength embedding on the multi-token embedding of the stellar labels.

\begin{figure*}
\centering
\resizebox{0.95\hsize}{!}
{\includegraphics[clip]{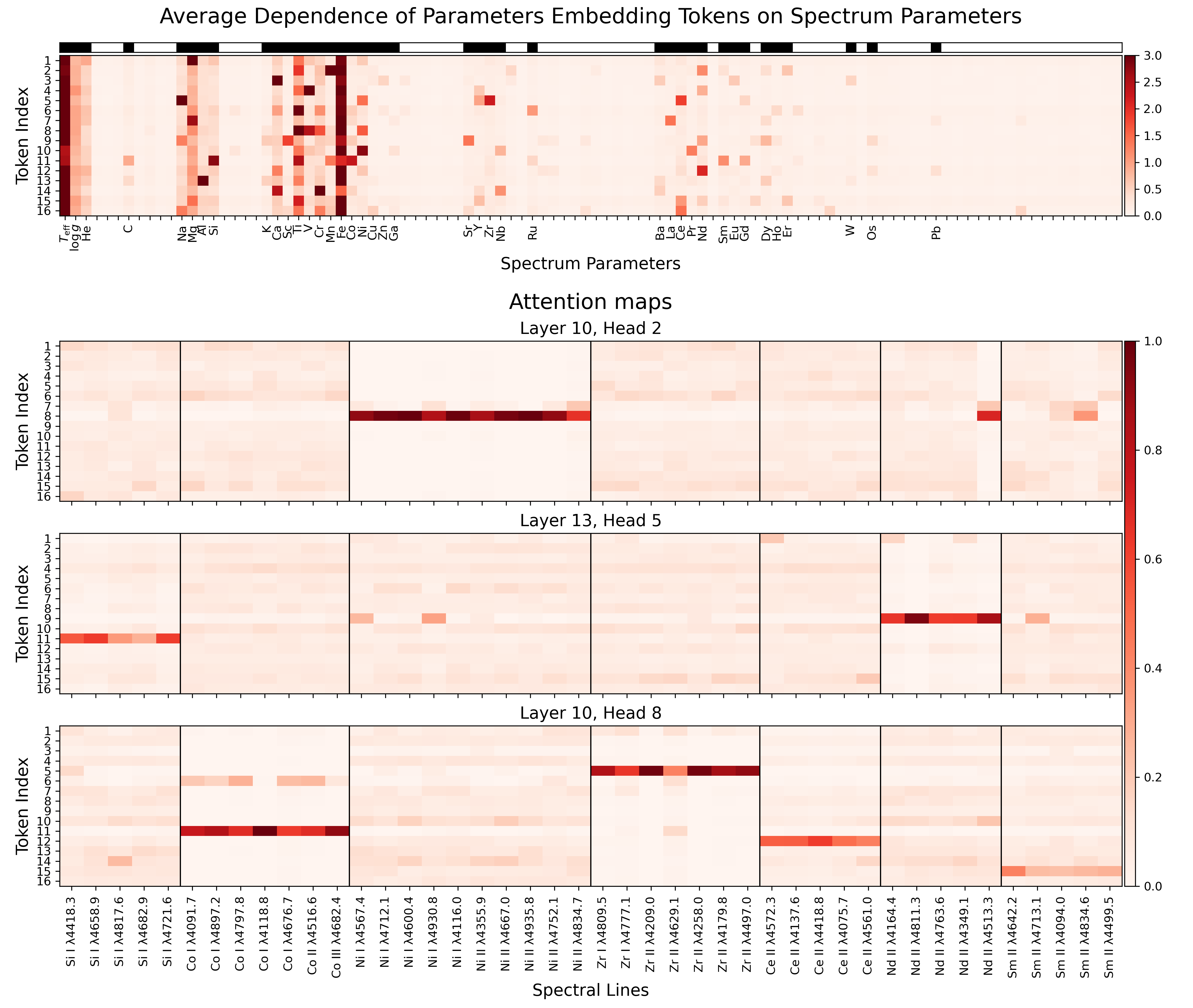}}
\caption{TransformerPayne extracts long-range information by assigning strong attention to wavelengths associated with the same elemental abundance. The upper panel of Fig. \ref{fig:fig9_interpretability} shows the Jacobian matrix $\partial \vec{p_{emb}} / \partial \vec{p}$, which represents the relationship between the parameter embedding $\vec{p_{emb}}$ and the corresponding spectrum parameters $\vec{p}$, averaged over the validation set. Most spectrum parameters are encoded in a subset of tokens. The lower panels showcase three representative attention maps, highlighting the attention between the embeddings of the wavelengths of spectral lines and their corresponding elements. TransformerPayne assigns strong attention to fluxes from multiple spectral lines of the same element, and the corresponding token index also corresponds to the respective element, as shown in the top panel. For instance, the attention map from layer 10, head 2, demonstrates prominent attention between the set of wavelengths corresponding to nickel absorption and the parameter token index 8, which is associated with the nickel element, as seen in the top panel. The same is observed for other elements, such as Si, Co, Zr, Ce, Nd, and Sm, as shown in the two bottom attention maps.}
\label{fig:fig9_interpretability}
\end{figure*}

Figure\,\ref{fig:fig9_interpretability} further demonstrates how TransformerPayne extracts long-range information. The upper panel shows the dependency of the parameters' embedding tokens, $\vec{p_{emb}}$, on the input parameters $\vec{p}$, by calculating the Jacobian matrices, $\partial \vec{p_{\text{emb}}} / \partial \vec{p}$, averaged over the validation dataset. If the token embedding depends on an input parameter, the associated Jacobian should be, on average, larger than the Jacobian with respect to other parameters. It is worth noting that since the embedding of the stellar labels is defined through an MLP network on the parameters, the parameters might not be encoded equally across all the tokens. Nonetheless, as shown, for the vast majority of the elements, the information of their abundances is encoded sparsely into a fixed token index. And the sparse encoding enables us to better interpret the attention as described in the following.

In the attention block of TransformerPayne, the tokenization of the wavelength ($\vec{w_{emb}}$) is cross-correlated with the parameters embedding, $\vec{p_{emb}}$, measured using the scalar product. Therefore, if TransformerPayne has learned the long-range information, we should expect the wavelength pertaining to a certain element to have a strong correlation with the corresponding element embedding. This is illustrated in the bottom panels of Fig.\,\ref{fig:fig9_interpretability}, which show the attention maps $\mathbf{A}_i$ (see equations in Sect.\,\ref{sec:TP_architecture}) from various layers of TransformerPayne for a chosen set of representative spectral lines of several elements and their corresponding attention values with the element. The attention maps demonstrate the expected "attention" as hypothesized.

For example, head 2 in layer 10 shows strong attention between the known Nickel lines and the particular token eight. The upper panels show that token eight is a token that primarily encodes the abundance of Nickel. The same behavior is also demonstrated across all elements. This shows that attention successfully learns to attend to the same tokens when predicting lines, even if they are widely separated in wavelength.

The strong attention observed in the results leads to the behavior we observe. On one hand, the effective inductive biases effectively reduce the degree of freedom of the emulators, allowing the training to be more efficient than The Payne (MLP-based models) as well as convolutional neural network models, and can achieve much more precise emulation. The training of such models is more computationally expensive and improves over the baselines after relatively long training, as shown in Fig.\,\ref{fig:fig6_scalling_steps}. This is because much of the first part of the training of models is spent searching for better tokenization, among other factors. However, once settled on the correct regime, the Transformer models lead to much more efficient and steeper improvement of the models.

On the other hand, the strong inductive biases also lead to the behavior where the emulation errors often show stronger correlation from pixels (Fig.\,\ref{fig:fig7_corr_matrix_residuals}) from the same elements, which can lead to biases in the inference. However, such bias is almost negligible for any practical purposes (Fig.\,\ref{fig:fig8_residuals_some_params}). Considering the much superior emulation precision, TransformerPayne is still the better option for emulation.

\subsection{Comparison with The Payne and Convolutional-based emulator}

The underperformance of convolutional neural networks (CNNs) in stellar spectrum emulation highlights a key challenge: applying convolutional networks to model the rapid changes observed in the stellar spectrum due to the presence of thousands of absorption features. It is challenging because convolutional neural networks are characterized by two inappropriate inductive biases. The first is translation invariance, which is the property of giving the same output for translated inputs. This property is often useful (like in image classification) but makes it difficult for a CNN-based neural network to learn the exact localization of particular features, especially since the localization of spectral lines in wavelengths is essential. The second is distortion stability, which means that the output of the CNN-based network is not particularly sensitive to mild distortions of input. This is useful when processing images or voice but not in the context of stellar spectra, where distortion of the line shapes carries relevant information and the abundance of some elements relies on very weak spectral features.

These inductive biases might be beneficial when integrating an MLP core with convolutional layers, as proposed in the so-called Convolutional-based emulator. However, this strategy requires careful consideration of the extent of up-sampling to apply through learned deconvolution. One relevant issue is that spectral lines vary from narrow (most lines of metals) to wide (e.g., hydrogen lines or molecular bands), so there is no single correct inductive bias (regarding the up-sampling part) for the entire spectrum. For this reason, the considered Convolutional-based emulator saturated with an MAE of emulation equal to 0.05. In summary, Convolutional-based emulators are not best suited for stellar spectra emulation, and The Payne-based approach might be a better alternative if TransformerPayne is too computationally heavy.

In contrast, The Payne, which uses the simplest fully connected network, might be a reasonable approach to deal with spectra. While the fully-connected network lacks inductive biases and is outperformed by TransformerPayne, it remains a viable approach for some cases. As we have shown here, massively scaling the complexity of the fully connected network can still continue to improve the emulation, albeit at a slower rate than TransformerPayne. The key advantage of The Payne is that the inference speed is much faster than TransformerPayne. Hence, when applied to vast samples, The Payne might still be the more feasible option. However, if higher prediction accuracy is required, especially with multiple elemental abundances, TransformerPayne shall take the lead role.

\subsection{Fine-tuning applied to stellar spectra emulation}

Fine-tuning is an effective strategy for stellar spectrum emulation. As demonstrated above, pre-training on fixed stellar parameters makes it simple for a model to learn the positions of the lines and their correlations due to changes in abundance. Then, the fine-tuning modifies the learned conditioning by extending it to include effective temperature and surface gravity dependence, without modifying the overall positions of spectral lines. The main effect of this strategy is that fewer spectra are needed to train the model to achieve the same emulation accuracy.

It was shown in detail in Sect.\,\ref{sec:fine_tuning}, that for example when targeting mean absolute error of emulation equal 0.005, TransformerPayne can be train with 4000 spectra when trained from scratch, but with 2000 spectra when using fine-tuning. This technique worked effectively for all considered emulators, but TransformerPayne has an added advantage because its inductive bias is particularly apt for this approach. This emulator can learn attention maps appropriate for a given wavelength range and reuse them during fine-tuning.

This approach has its parallels in large language models, where it is, for example, observed that learning to code in one programming language correlates with results in programming in another programming language \citep{2023arXiv230812950R}. This parallel shows that learning one task might enable the model to easily generalize to another associated task.

\subsection{Toward few-shot learning for spectral emulation}

Few-shot learning refers to the capability of a machine learning model to adapt to a new task using only a few examples (e.g., up to around 10). This capability is particularly relevant in the context of stellar spectra emulation. For instance, consider the spectra obtained from the JWST NIRSpec instrument. With few-shot learning for stellar spectra, we could use a dozen spectra of standard stars to calibrate the emulator for accurate inference of stellar atmospheric parameters from JWST spectra. Few-shot learning is essential in this context due to the lack of large-scale observational datasets for these wavelengths.

TransformerPayne is a promising architecture to demonstrate few-shot capabilities when scaled to larger networks and training datasets. This potential arises from the sparse representations it learns internally, as illustrated in Fig.\,\ref{fig:fig9_interpretability}. It shows that TransformerPayne attends to the same tokens for distant wavelengths, encouraging more predictive internal features. When adapting to a new domain, the features need modification, but not necessarily the attention maps.

In the context of sparse and interpretable attention maps, few-shot learning can be understood as follows: without shared internal features, each spectral line prediction would rely on its own internal feature, resulting in millions of independent features. However, if internal features are shared between lines, their number is significantly reduced. In this case, when adapting to a new domain, the relevant features are updated using many more observations, equivalent to the number of spectra times the number of relevant lines. In short, shared and sparse features are much more likely to be well constrained from just a few spectra. Scaling is a crucial component because it is typically associated with more interpretable and simpler internal features.

Regarding future work that is a direct application of our findings, one possible direction is to study the feasibility of few-shot learning in the context of precise inference of individual abundances with 3D non-LTE modeling. 3D non-LTE modeling is prohibitively computationally expensive for the calculation of large spectra grids. Application of the TransformerPayne emulator together with transfer learning might potentially make it feasible within current computational capabilities.

\subsection{Limitation and future work}

The main limitation of Transformer Payne is its decreased inference speed. When compared to the BigPayne emulator, the prediction of a single spectrum using TransformerPayne is about 30 times slower on a laptop using a CPU (3 seconds and 0.1 seconds, respectively) and 80 times slower on a GPU (30 milliseconds and 0.35 milliseconds, respectively). The measured speeds of The Payne and BigPayne are comparable. The TransformerPayne is significantly slower when compared to The Payne and BigPayne, as it involves many layers of Multi-Head Attention and Feed Forward networks, which are processed sequentially, whereas The Payne contains only several layers of matrix multiplication. Despite the relative inefficiency of TransformerPayne, it is significantly faster than traditional spectrum synthesis methods, which can take several minutes.

There are several approaches to improve the speed of the prediction of TransformerPayne, e.g., usage of efficient implementation of attention operation called Flash Attention \citep{dao2023flashattention2} or quantization of model's weights to lower precision \citep{kim2023stack}. Flash Attention optimizes the attention mechanism by implementing it with low-level CUDA kernels and optimizing memory management. Quantization replaces the model weights with lower precision approximations, which modern GPUs are optimized for.

\section{Conclusions}
\label{sec:conclusions}

Emulating spectra is critical for precise inference of parameters of stellar atmospheres. However, the current state-of-the-art emulator, The Payne, tends to saturate in emulation accuracy and requires a relatively large number of spectra. We hypothesize that this is caused by the small scale of The Payne emulator and the lack of appropriate inductive biases of its architecture.

To overcome these limitations, we introduce TransformerPayne in this study. We leverage the attention mechanism to harness the long-range information in the spectra. In particular, we propose a wavelength parametrization that allows the attention mechanism to effectively capture the dependence between wavelengths with vast separation and associate them with the corresponding elements.

We found that the TransformerPayne architecture performs state-of-the-art spectral emulation quality. In terms of emulation error in reproducing the fluxes of the spectrum, TransformerPayne typically outperforms The Payne tenfold and BigPayne, the massively scaled-up version of The Payne, twofold. We also found that convolutional neural network-based emulators, with their inadequate inductive biases, tend to perform much worse than even the vanilla MLP models of The Payne. Our study shows that convolutional neural networks might not be an adequate architecture to deal with spectra effectively because of their inappropriate inductive biases: translation invariance and distortion stability. Even when chained together with MLPs, as in our case, they are not comparable to pure MLP networks like The Payne and BigPayne.

The superior performance in emulation from TransformerPayne further translates into more accurate label recoveries. By fitting our validation models, we found that the typical recovery error from the emulation errors is minimal for TransformerPayne. Assuming a wavelength range from 4000\,\r{A} to 5000\,\r{A} and a resolution of 100\,000, the typical abundance errors from TransformerPayne range from 0.03 dex for some of the weakest elements with limited spectral features to 0.004 dex for the more prominent elements such as iron. This is 10 times more precise than the classical The Payne with limited expressivity and remains 1-2 times more precise than the scaled-up version of The Payne.

As the Transformer architecture is well-suited to deal with spectral emulation, TransformerPayne also demonstrates a much steeper improvement in emulation error with respect to the number of training steps and shows no sign of saturation in performance up to a training set of 100\,000. This demonstrates that TransformerPayne is a scalable model and can continue to improve with sufficient computational resources.

Furthermore, due to its inherent ability, TransformerPayne benefits from the pretraining-fine-tuning strategy. The complex spectral emulation task benefits from a more fine-grained divide-and-conquer strategy, first pre-training on the pre-training grid with the same stellar parameters and only varying the elements to facilitate the attention mechanism to learn the correlation between the pixels. Then, the model is fine-tuned on a larger grid with varying stellar parameters as well. This allows one to decrease the need for a large training set for the latter by almost an order of magnitude. This method paves the way effectively training expensive synthetic models, such as 3D NLTE models, with only hundreds of training spectra.

TransformerPayne also exhibits better interpretability. We showed that the learned attention of the models exhibits sparsity, where attention is given to wavelength pixels from the same underlying elemental abundances. Such a sparse representation is a tell-tale sign of a generalizable model for fine-tuning from a small sample. This enables a new path toward better calibrating spectral models as well as understanding missing spectral features.

While the current models of TransformerPayne still suffer from slow inference speed, which might require some further engineering and architectural optimizations to be vastly deployed for pipelines for spectroscopic surveys, the TransformerPayne architecture marks an advancement in the topic of stellar spectra emulation. It offers improved accuracy, data efficiency, and interpretability over existing spectral emulation methods, resolving one of the critical challenges presented in the analysis of spectral data.

\begin{acknowledgments}
YST acknowledges financial support received from the Australian Research Council via the DECRA Fellowship, grant number DE220101520. This research was undertaken with the assistance of resources and services from the National Computational Infrastructure (NCI), which is supported by the Australian Government. TR and YST gratefully acknowledge the financial support received from the University of Chicago Data Science Institute for an extensive visit. Additionally, we extend our heartfelt thanks to Alex Ji, David Weinberg, Jennifer Johnson, Anil Pradhan, Adam Wheeler, Anish Amarsi, Jiří Kubát, Ewa Niemczura, Maria Bergemann, Nicholas Storm, Richard Hoppe and Philip Eitner for invaluable discussions.
\end{acknowledgments}

\vspace{5mm}

\software{matplotlib \citep{Hunter_2007},
          jax \citep{jax2018github},
          flax \citep{flax2020github},
          optax \citep{deepmind2020jax}
          }

\appendix

\section{Architecture details}
\label{arch_details}

The code snippets below demonstrate the Python implementation of tested emulator architectures, using the \texttt{jax} and \texttt{flax} libraries, with adjustments made for enhanced readability. Note that the TransformerPayne code predicts a scalar value representing the normalized flux at a given wavelength, which then needs to be vectorized, whereas the other models predict fluxes across fixed sets of wavelengths.

\begin{lstlisting}[language=Python, caption=The Payne code.]
class thePayne(nn.Module):
  @nn.compact
  def __call__(self, labels, train):
    _x = labels
    for features in (128, 128):
      _x = nn.Dense(features)(_x)
      _x = nn.gelu(_x)
    _x = nn.Dense(22135)(_x)
    return nn.sigmoid(_x)
\end{lstlisting}

\begin{lstlisting}[language=Python, caption=The BigPayne code.]
class theBigPayne(nn.Module):
  @nn.compact
  def __call__(self, labels, train):
    _x = labels
    for features in (2048, 2048, 2048):
      _x = nn.Dense(features)(_x)
      _x = nn.gelu(_x)
    _x = nn.Dense(22135)(_x)
    return nn.sigmoid(_x)
\end{lstlisting}

\begin{lstlisting}[language=Python, caption={The Convolutional-based emulator code.}, label={lst:simplified_convnn}]
class Convolutional_based(nn.Module): 
  @nn.compact
  def __call__(self, inputs, train):
    _x = inputs
    for features in (2048, 2048, 2048):
      _x = nn.Dense(features)(_x)
      _x = nn.gelu(_x)

    _x = jnp.reshape(_x, (512, 4))
    
    for _ in range(5):
      _x = nn.ConvTranspose(features=8, 
                kernel_size=(4,), 
                strides=(2,), 
                padding='SAME')(_x)
      _x = nn.gelu(_x)
      
      _x = nn.Conv(features=32, 
                kernel_size=(3,), 
                strides=(1,), 
                padding='SAME')(_x)
      _x = nn.gelu(_x)
      
      _x = nn.Conv(features=32, 
                kernel_size=(3,), 
                strides=(1,), 
                padding='SAME')(_x)
      _x = nn.gelu(_x)

    _x = nn.Conv(features=1, 
                kernel_size=(1,), 
                padding='SAME')(_x)
    _x = jnp.reshape(_x, (-1,))
    
    w_in = jnp.linspace(0, 1, 16384)
    w_out = jnp.linspace(0, 1, 22135)
    _x = jnp.interp(w_out, w_in, _x)

    return nn.sigmoid(_x)
\end{lstlisting}

\begin{lstlisting}[language=Python, caption={The TransformerPayne code.}, label={lst:transformer_payne}]
from flax.linen import \
  MultiHeadDotProductAttention as MHA

def frequency_encoding(x, 
                  min_period, 
                  max_period, 
                  dim):
  lp0 = jnp.log10(min_period)
  lp1 = jnp.log10(max_period)
  p = jnp.logspace(lp0, lp1, num=dim)
  
  return jnp.sin(2 * jnp.pi / p * x)

class TransformerPayne(nn.Module):
  @nn.compact
  def __call__(self, x):
    p, w = x
    enc_w = frequency_encoding(w, 
                        1e-6, 
                        10, 
                        256)[None, ...]
                        
    p = nn.Dense(4*256)(p)
    p = nn.gelu(p)
    p = nn.Dense(16*256)(p)
    p = jnp.reshape(p, (16, 256))
    enc_p = nn.LayerNorm()(p)
    
    x_pre = x_post = enc_w
    for _ in range(16):
      attn_layer = MHA(num_heads=8)
      _x = attn_layer(inputs_q=x_post, 
                      inputs_kv=enc_p)
      _x = _x + x_post
      x_post = nn.LayerNorm()(_x)
      x_pre += _x
      
      _x = nn.Dense(4*256)(x_post)
      _x = nn.gelu(_x)
      _x = nn.Dense(256)(_x)
      _x = _x + x_post
      x_post = nn.LayerNorm()(_x)
      x_pre += _x
    
    norm = nn.LayerNorm()(x_pre + x_post)
    x = nn.Dense(256)(norm[0])
    x = nn.gelu(x)
    
    return nn.Dense(1)(x)
\end{lstlisting}


\bibliography{references}{}
\bibliographystyle{aasjournal}



\end{document}